\pdfoutput=1
\documentclass[conference]{IEEEtran}
\ifCLASSINFOpdf
\else
\fi
\usepackage[cmex10]{amsmath}
\usepackage{url}

\usepackage{graphicx} 
\usepackage{makecell}
\usepackage{subfigure}
\usepackage[sort, numbers]{natbib}%

\makeatletter
\IEEEtriggercmd{\reset@font\tiny\fontsize{5.2pt}{5.5pt}\selectfont}
\makeatother
\IEEEtriggeratref{1}

\makeatletter \g@addto@macro{\UrlBreaks}{\do\/\do\-} \makeatother

\let\sigproof\proof\let\proof\relax
\let\sigendproof\endproof\let\endproof\relax
\usepackage{amsthm}
\usepackage{algorithmic}
\renewcommand{\algorithmicrequire}{\textbf{Input:}}
\renewcommand{\algorithmicensure}{\textbf{Output:}}
\usepackage[group-separator = {,}]{siunitx}

\usepackage[ruled]{algorithm2e}

\SetAlFnt{\small}
\SetAlCapFnt{\small}
\SetAlCapNameFnt{\small}
\SetAlCapHSkip{0pt}
\IncMargin{-\parindent}

\let\proof\sigproof
\let\endproof\sigendproof

\newtheoremstyle{sig}
  {}
  {}
  {\itshape}
  {}
  {\scshape}
  {.}
  {.5em}
  {#1 #2\thmnote{\quad(#3)}}

\makeatletter

\newcommand{\Rmnum}[1]{\expandafter\@slowromancap\romannumeral #1@}

\theoremstyle{sig}
\newtheorem{thm}{Theorem}[section] 
\newtheorem{defn}[thm]{Definition} 
\newtheorem{rmk}[thm]{Remark} 

\newcommand{\pnum}[1]{\SI{#1}{\percent}}
\newcommand{\cnum}[1]{\num[group-separator = {,}, group-minimum-digits = 4]{#1}}

\begin{document}
%
\title{Smoke Screener or Straight Shooter: Detecting Elite Sybil Attacks in User-Review Social Networks}

\author{\IEEEauthorblockN{Haizhong Zheng}
\IEEEauthorblockA{Shanghai Jiao Tong University\\
sjtu.zhenghaizhong@gmail.com}
\and
\IEEEauthorblockN{Minhui Xue}
\IEEEauthorblockA{NYU Shanghai and ECNU\\
minhuixue@gmail.com}
\and
\IEEEauthorblockN{Hao Lu}
\IEEEauthorblockA{Shanghai Jiao Tong University\\
luhao0522@gmail.com}
\and
\IEEEauthorblockN{Shuang Hao}
\IEEEauthorblockA{University of Texas at Dallas\\
shao@utdallas.edu}
\and
\IEEEauthorblockN{Haojin Zhu}
\IEEEauthorblockA{Shanghai Jiao Tong University\\
zhu-hj@cs.sjtu.edu.cn}
\and
\IEEEauthorblockN{Xiaohui Liang}
\IEEEauthorblockA{University of Massachusetts Boston\\
xiaohui.liang@umb.edu}
\and
\IEEEauthorblockN{Keith Ross}
\IEEEauthorblockA{NYU and NYU Shanghai\\
keithwross@nyu.edu}
}


%


\IEEEoverridecommandlockouts
\makeatletter\def\@IEEEpubidpullup{9\baselineskip}\makeatother
\IEEEpubid{\parbox{\columnwidth}{
    Network and Distributed Systems Security (NDSS) Symposium 2018\\
    18-21 February 2018, San Diego, CA, USA\\
    ISBN 1-1891562-49-5\\
    http://dx.doi.org/10.14722/ndss.2018.23009\\
    www.ndss-symposium.org%
}
\hspace{\columnsep}\makebox[\columnwidth]{}}


\maketitle

\begin{abstract}
Popular User-Review Social Networks (URSNs)---such as Dianping, Yelp, and Amazon---are often the targets of reputation attacks in which fake reviews are posted in order to boost or diminish the ratings of listed products and services. These attacks often emanate from a collection of accounts, called Sybils, which are collectively managed by a group of real users. A new advanced scheme, which we term elite Sybil attacks, recruits organically highly-rated accounts to generate seemingly-trustworthy and realistic-looking reviews. These elite Sybil accounts taken together form a large-scale sparsely-knit Sybil network for which existing Sybil fake-review defense systems are unlikely to succeed.

In this paper, we conduct the first study to define, characterize, and detect elite Sybil attacks. We show that contemporary elite Sybil attacks have a hybrid architecture, with the first tier recruiting elite Sybil workers and distributing tasks by Sybil organizers, and with the second tier posting fake reviews for profit by elite Sybil workers. We design \textsc{ElsieDet}, a three-stage Sybil detection scheme, which first separates out suspicious groups of users, then identifies the campaign windows, and finally identifies elite Sybil users participating in the campaigns. We perform a large-scale empirical study on ten million reviews from Dianping, by far the most popular URSN service in China. Our results show that reviews from elite Sybil users are more spread out temporally, craft more convincing reviews, and have higher filter bypass rates. We also measure the impact of Sybil campaigns on various industries (such as cinemas, hotels, restaurants) as well as chain stores, and demonstrate that monitoring elite Sybil users over time can provide valuable early alerts against Sybil campaigns.
\end{abstract}

\section{Introduction}

User-Review Social Networks (URSNs)---such as Dianping, Yelp, and Amazon---are often the targets of Sybil attacks, where multiple fake accounts, called Sybils, are used to generate fake reviews that masquerade as testimonials from ordinary people. The goal of the attack is to deceive ordinary users into making decisions favorable to the attackers.
A recent evolutionary trend is a new type of Sybil attack in contemporary URSNs, which we call \emph{elite Sybil attacks}. Elite Sybil attacks recruit highly-rated users (\textit{e.g.},  ``Elite'' member on Yelp or ``5-star'' member on Dianping) who normally post genuine reviews, unbiased by financial incentives.
Directed by organizational leaders, elite Sybil attackers mimic the behavior of real users by posting topically coherent content with temporal patterns consistent with real users.
Because elite Sybil users' review behavior greatly resembles that of genuine users, elite Sybil attacks are extremely difficult to algorithmically or manually detect. Therefore, new approaches are needed to detect elite Sybil accounts rapidly and accurately.

\noindent \textbf{Challenges.} Previous work on defending against Sybil attacks in Online Social Networks (OSNs) aims to identify fake or compromised accounts mainly by two means: (i) investigating an account's social network connectivity~\cite{viswanath2010analysis, cao2012aiding, yu2008sybillimit, yu2006sybilguard, hu2013social} relying on the trust that is established in existing social connections between users; (ii) building machine learning classifiers with a set of identified features~\cite{egele2013compa, stein2011facebook, zhao2009botgraph}. The literature on Sybil defense schemes mostly targets general OSNs, and almost no reasons are tailored toward a situational logic behind that attack, much less pay attention to Sybil defenses in URSNs, such as Yelp and Dianping. URSNs pose the following three unique challenges. (i) The nodes in URSNs do not exhibit tight connectivity as in general OSNs, rendering graph-connectivity based approaches less effective in URSNs. (ii) Elite Sybil attacks in URSNs are more professional, writing elaborate reviews and posting related pictures to imitate real reviews. Thus, Sybil attacks in URSNs are more difficult to detect than those in traditional OSNs. (iii) Since elite Sybil attackers only contribute to a small fraction of overall reviews, the existing Sybil detection approaches based on the similarity of aggregate behavior do not work well.
To address all these challenges and deficiencies, a novel Sybil detection technique for elite Sybil users is highly desired.


\noindent \textbf{\textsc{ElsieDet}.} In this work, we design a novel {\bf El}ite {\bf Sy}bil {\bf Det}ection system, \textsc{ElsieDet}, which can identify URSN Sybil users with elaborate camouflage. Different from previous studies, we focus our design on Sybil campaigns that have multiple Sybil workers colluding to perform a task (\textit{e.g.}, posting positive reviews and high ratings for a specific restaurant) under the coordination of a Sybil leader. These campaigns have an active time period. Any user who posts during the active time period is suspicious to be part of the campaign. This user could either be a benign user who happens to visit the store and post her review in the campaign period, or a Sybil user who posts fake reviews specifically for the campaign. We build \textsc{ElsieDet} based on the following empirical observations: A benign user posts honest reviews based on her real experience while a Sybil user always posts fake reviews during the active time period of the Sybil campaigns. Therefore, in the long run, the more campaigns a user gets involved in, the more likely she is a Sybil user. 

\textsc{ElsieDet} is designed with three stages: detecting a Sybil community (Phase~I), determining the Sybil campaign time window (Phase~II), and finally classifying elite Sybil users (Phase~III). In Phase~I, since Sybil users collaborate to post fake reviews in a Sybil campaign, \textsc{ElsieDet} exploits this group behavior to cluster users and identify Sybil communities. In Phase~II, \textsc{ElsieDet} uses a novel campaign detection algorithm to automatically determine the start and end points of a Sybil campaign, while ruling out reviews not belonging to a Sybil task. Lastly, in Phase~III, we propose a novel elite Sybil detection algorithm to separate out elite Sybil users from undetected users based on a new defined metric, \emph{Sybilness}, which scores the extent a user participates in the Sybil campaign.

We implement \textsc{ElsieDet} and evaluate its performance on a large-scale dataset from Dianping, by far the most popular URSN in China. Our dataset was crawled from January 1, 2014 to June 15, 2015 and includes $10,541,931$ reviews, $32,940$ stores, and $3,555,154$ users. We show that, of all the reviews, more than $108,100$ reviews are fake reviews, which were generated by $21,871$ regular Sybil users and $12, 292$ elite Sybil users. These Sybil users belong to $566$ Sybil communities, which launched $2,164$ Sybil campaigns. Our research shows that the current filtering system of Dianping is ineffective at detecting fake reviews generated by the elite Sybil users since less than $33.7\%$ of the fake reviews have been filtered by Dianping. Finally, through manual inspection, we conclude that $90.7\%$ of randomly sampled suspicious users are elite Sybil users, and $93.8\%$ of the $1,000$ most suspicious users are elite Sybil users. We have reported all of our findings to Dianping, which acknowledged our detection results.

{\bf \noindent Findings.} Our study reveals the following main findings about the operation logistics of elite Sybil attacks. 
\begin{itemize}
\item Motivated by economic revenue on black markets (\textit{e.g.}, an elite Sybil user  can receive up to $20$ times more income than a regular Sybil user for the same task), elite Sybil users have developed a series of techniques  to evade the Sybil detection systems, including coordinating the posting time and crafting carefully-polished review contents and pictures.

\item We evaluate the impact of Sybil attacks on different categories of industry. Surprisingly, cinemas, hotels, and restaurants are the most active in hiring Sybil users for promotions. In particular, $30.2\%$ of cinemas, $7.7\%$ of hotels, and $5.5\%$ of restaurants are actively involved in Sybil campaigns.

\item We observe that $12.4\%$ of Sybil communities post fake reviews for chain stores, which is different from recent research performed on Yelp~\cite{harvard2016fake}. What is more interesting is that that overhyped chain stores with the same brand recruit the same Sybil communities for Sybil campaigns.

\item We find that more than $50\%$ of Sybil campaigns can be determined within the first two weeks by only observing activities of elite Sybil users, thereby allowing the URSN to defend against the attack while in progress.
\end{itemize}

\noindent \textbf{Contributions.}
To the best of our knowledge, our work is the first to study elite Sybil detection in URSNs. In summary, we make the following key contributions:
\begin{enumerate}
\item We show that the Sybil organization of Dianping has evolved to a hybrid architecture, rather than a prevalent centralized or a simple distributed system~\cite{wang2012serf, song2015crowdtarget}.
\item We identify a new type of Sybil users, elite Sybil users, which employ a sophisticated strategy for evading detection and have never been studied before. 
\item We characterize the behaviors of elite Sybil users and propose an early-warning system to detect online Sybil campaigns.
\item We show that \textsc{ElsieDet} complements the Dianping's current filtering system, which has been verified by both our own manual inspection and the feedback received from Dianping. 
\end{enumerate}

\noindent \textbf{Roadmap.} The remainder of this paper is structured as follows: Section~\ref{background} introduces the necessary background on Dianping and Sybil attacks while Section~\ref{ssec: compare crowturfing} defines elite Sybil attacks. In Section~\ref{design_implementation}, we propose our Sybil detection system. Section~\ref{section:evaluation} evaluates the experimental performance, whereas Section~\ref{measurement} provides detailed measurements of elite Sybil users and Sybil communities. Section~\ref{ssec:application} discusses applications and limitations of the study. Section~\ref{relatedwork} surveys the related work. Finally, Section~\ref{conclusion} concludes the paper.

\subsection{Ethical Considerations}

In this paper, we only collected publicly available review information and its relation with stores on Dianping. We do not crawl, store, or process users' privacy information including usernames, gender, small profile pictures, or tags that often accompany the user profiles. Furthermore, we did not craft fake reviews in order to ensure that our experiments do not have a negative impact on Dianping's services. Finally, we have alerted Dianping about the discoveries and results made in this paper. We are currently discussing possibilities of our system deployment at Dianping.

%

\section{Background}\label{background}
In this section, we first briefly describe Dianping. We then summarize traditional Sybil attacks and the recent trend on User-Review Social Networks (URSNs).

\subsection{Dianping: A User-Review Social Network}\label{ssec:Dianping}


Dianping is by far the most popular URSN in China, where users can review local businesses such as restaurants, hotels, and stores. 
When a user uses Dianping, Dianping will return to the user with a list of choices in order of overall quality-rating.
The quality-rating of a restaurant review is typically scaled from $1$ star (worst) to $5$ star (best), mainly depending on the restaurant service.
Users are also assigned star-ratings. These star-ratings vary from $0$ stars (rookie) to $6$ stars (expert), depending on the longevity of the user account, the number of reviews posted, etc.
A higher star-rating indicates that the user is more experienced and more likely to be perceived as an expert reviewer. Similar to ``Elite User'' on Yelp, a senior level user (\textit{e.g.}, 4-star, 5-star, or 6-star user) is supposed to be a small group of in-the-know users who have a large impact on their local community. Dianping has established its user reputation system that classifies user reviews into ``normal reviews'' and ``filtered reviews.'' The latter includes the uninformative reviews or the suspicious reviews that are potentially manipulated by the Sybil attackers, but the details of the algorithm remain unknown to the public.

\subsection{Sybil Attacks}
Social media platforms populated by millions of users present either economic or political incentives to develop algorithms to emulate and possibly alter human behavior.  
Earlier Sybil attacks include malicious entities designed particularly with the purpose to harm. These Sybil users mislead, exploit, and manipulate social media discourse with rumors, spam, malware, misinformation, slander, or even just noise~\cite{heymann2007fighting, gupta20131}. 
This type of abuse has also been observed during the 2016 US presidential election~\cite{allcott2017social}.
As better detection systems are built, we witness an arms race similar to what we observed for spam alike in the past. In recent years, Twitter Sybils have become increasingly sophisticated, making their detection more difficult. For example, Sybils can post collected material searched from websites at predetermined times, emulating the human temporal signature of content production and consumption~\cite{golder2011diurnal}. 
In the meantime, the arms race has also driven the corresponding countermeasures~\cite{egele2013compa,CCS14syn,CCS14twitter,NDSS2015integro}.

The evolutionary chain of Sybil attacks imposes a novel challenge in the most-up-to-date URSNs: They provide fake content among little pieces of their information, regardless of their accuracy, which is highly popular and endorsed by many high-level organizers, exerting great impact against which there are no effective countermeasures.
In this paper, we characterize and detect a new type of Sybil attacks in URSNs, typically applying our methodology to Dianping as our case study.

\section{Dissecting Elite Sybil Attacks}\label{ssec: compare crowturfing}
In this section, we first introduce some definitions. We then define a novel type of Sybil attackers, coined as \emph{elite Sybil users}. We finally take an in-depth dive into the typical hierarchical architecture and the key actors playing in a Sybil organization.

\subsection{Terminology}\label{definitions} 
To formulate our problem precisely, we introduce the following definitions.

\begin{defn}
\noindent \textbf{Store:} A Store $S$ has an official website on Dianping that contains a large number of reviews of this particular store.
\end{defn}

\begin{defn}
\noindent \textbf{Community}: A Community $C$ is a group of users who post reviews in similar stores to rate and comment such stores.
\end{defn}

\begin{rmk}
In our paper, we categorize all communities into two types: \textbf{benign communities} and \textbf{Sybil communities}. We define a benign community to be formulated by all benign (real, normal) users and a Sybil community to be formulated by all Sybil (malicious) users. A set of users is also partitioned into two types: A \textbf{benign user} is a person who posts honest reviews and a \textbf{Sybil user} is a person who posts fake reviews to boost the prestige of stores. We will use the terms benign users and real users interchangeably.
\end{rmk}

\begin{defn}
\noindent \textbf{Campaign:} A campaign---denoted as $(C, S, T_s, T_e)$, where $C, S, T_s, T_e$ denote community ID, store ID, starting time, and ending time of a campaign---is an activity in which users of a Community $C$ post reviews in Store $S$ from $T_s$ to $T_e$ to boost the prestige of Store $S$.
\end{defn}

\begin{rmk}
For Sybil users in a given community, these Sybil users serve for various stores. Each of these stores has one particular campaign launched by this community. However, these stores can have other campaigns, but are launched by other communities.
\end{rmk}

\subsection{Elite Sybil Users}\label{elite}
In a Sybil organization of Dianping, we find a new type of Sybil users, termed \emph{elite Sybil users}. Different from \emph{regular Sybil users} studied before, elite Sybil users post reviews not belonging to Sybil tasks, which can harm the accuracy of existing detection systems to a large degree. Elite Sybil accounts are mainly composed of two kinds of accounts: either
(i) Sybil accounts created reviews not belonging to Sybil tasks (smoke-screening) in order to mimic genuine users purely for the use of campaigns; or (ii) accounts owned by benign users---usually with high-rating stars---that convert to Sybil accounts when fulfilling a Sybil task within a campaign in order to reap the rewards offered by Sybil organizations (The Sybil task is detailed in Section~\ref{workflow}.).
Although elite Sybil accounts belong to multiple users/entities, they are manipulated by a single entity (\emph{i.e.}, Sybil leader). This satisfies the definition of Sybil attack that a malicious entity takes on multiple identities. Therefore we consider the attack performed by elite Sybil accounts as Sybil attack.
By hiding behind massive reasonable reviews posted however deliberately or unwittingly, these reviews posted by elite Sybil users appear realistic as those posted by benign users. Compared with regular Sybil users, elite Sybil users are more active out of the Sybil campaigns, which enables elite Sybil users to have a much lower percentage of fake reviews in their posts and higher user-level star-ratings (see Section~\ref{measurement}).

\noindent {\bf Black market and economic factors.} Here, we try to explore the monetary reward for an elite Sybil user on Dianping. Table~\ref{Table:fee for Sybil worker} shows hierarchical rewards for a specific Sybil organization into which we infiltrated recently. We see that the rewards depend on the ratings of Sybil accounts. Not surprisingly, the monetary rewards earned by each \emph{submission} increase as the ratings of accounts increase. This is largely because users with higher ratings have a larger influence, their reviews are less likely to be deleted, and thus are more attractive to Sybil organizers. Likewise, the reviews from the highly-ranked users are more influential, and have a larger chance of being presented in the front page of a store, which can potentially attract more attention from customers.

\begin{table}[!ht]\small
  \caption{Hierarchical rewards for (elite) Sybil workers}
 \label{Table:fee for Sybil worker}
 \centering
 \begin{tabular}{|c|c|}
 \hline {\bf  Ratings of Accounts}  & {\bf Rewards per Submission}  \\
 \hline
 \hline 0-star, 1-star	  	& \$0.30\\
  \hline 2-star  		& \$0.75 \\
  \hline 3-star 	   	& \$1.50  \\
  \hline 4-star  	        	& \$3.74 \\
  \hline 5-star, 6-star	        	& {\bf \$5.98}  \\
  \hline
 \end{tabular}
\vspace{-0.2cm}
\end{table}



\subsection{Anatomy of Elite Sybil Attack Operations} \label{ssec:sybil-attack}

Many review websites are suffering from review manipulation, which can be seen as a variant of Sybil attacks in URSNs. Similar to Yelp and TripAdvisor, Dianping is struggling with review manipulation as well. To investigate these organizations in depth, we impersonated Sybil users in order to investigate how the tasks are distributed and executed by the Sybil organizer. Note that, for the ethical considerations, we did not perform any real tasks in reality.  In most cases, Sybil leaders regularly post contact information on social media (\emph{e.g.}, Tencent QQ, WeChat, and Douban Group\footnote{Douban Group, being part of Douban, is composed of huge numbers of sub forums for users to post messages under various topics. \url{https://www.douban.com/group}}) to attract Sybil workers. Specifically, we acquired contact information from Douban Group to reach out many Sybil organizations. During our month-long investigation, we found that the Sybil attacks on Dianping show a unique organization pattern.

\begin{figure}[t]
  \begin{center}
  \includegraphics[width = 0.4 \textwidth]{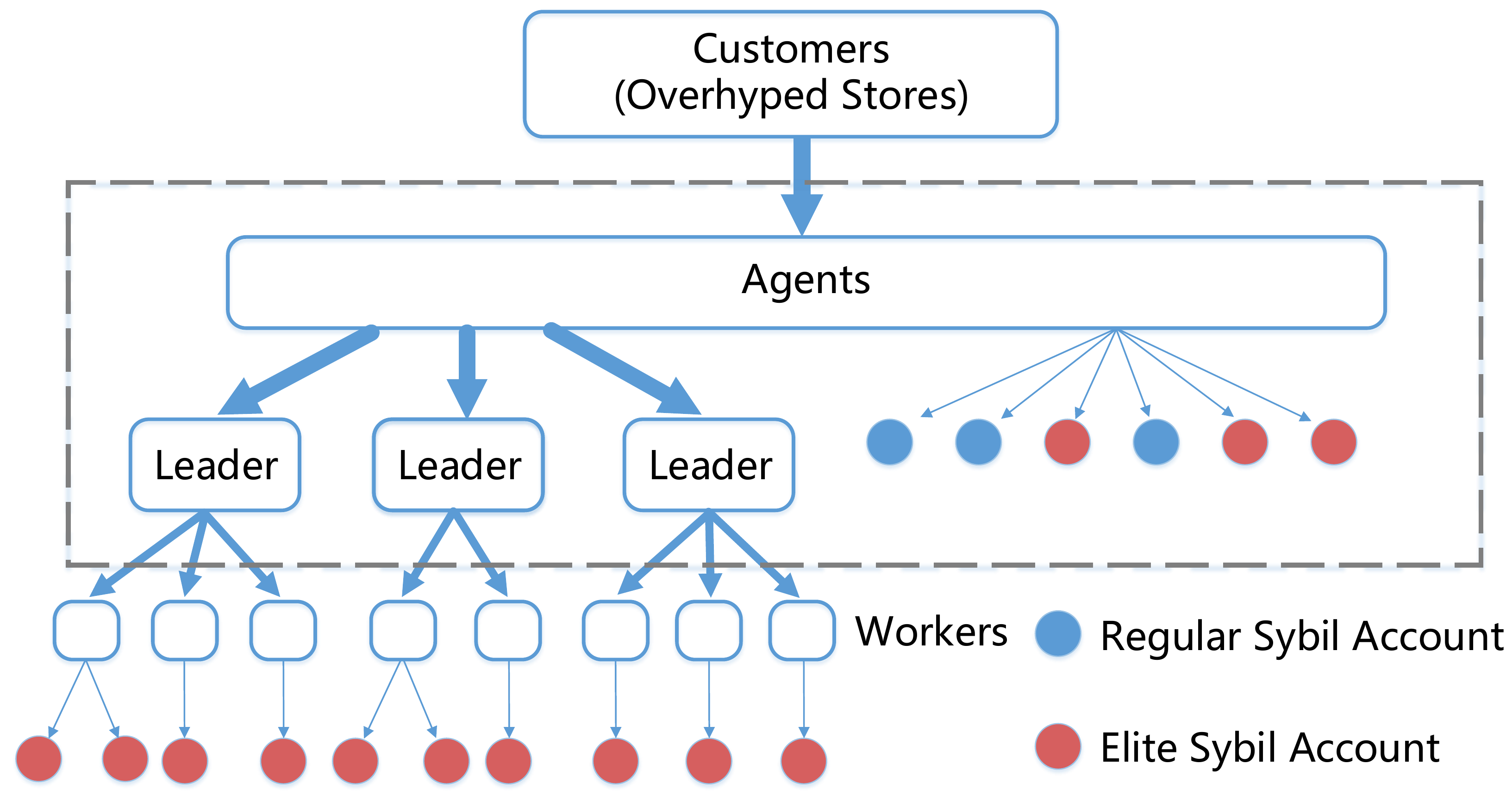}
  \caption{The architecture of a Sybil organization}\label{Fig:arhcitecture of sybil organization}
  \end{center}
  \vspace{-0.6cm}
\end{figure}

\subsubsection{A Hybrid Architecture}\label{Sybil Organization}
Sybil organizations usually show either a centralized or distributed architecture on Facebook or Twitter. The Sybil organization on Dianping, however, evolves to a hybrid architecture, which involves four key actors, as shown in Figure~\ref{Fig:arhcitecture of sybil organization}:
\begin{enumerate}
\item \emph{Customers (or Overhyped stores):} Businesses that want to boost their scores rapidly on Dianping. Overhyped stores propose mission description and monetary rewards for a Sybil organization to launch Sybil campaigns. They are beneficiaries from Sybil campaigns.
\item \emph{Agents:} Organizers are agents who are responsible for accepting the tasks from overhyped stores and upper management of a Sybil organization. Organizers take charge of launching the Sybil campaigns.
\item \emph{Leaders:} Leaders take charge of recruiting Sybil workers and make arrangements for crafting reviews. Leaders distribute tasks to Sybil workers and payment.
\item \emph{Elite Sybil workers:} Elite Sybil workers are Internet users, recruited by leaders, who post fake reviews for profit. These elite Sybil accounts are then manipulated by elite Sybil workers to post fake reviews. (Elite Sybil accounts, users, and workers are interchangeable in this paper.)
\end{enumerate}

In this architecture, the leader plays a key role in task distribution and quality control of review comments for the following reasons: First, the leader himself/herself controls a certain number of Sybil accounts, and these facilitate the launch of a campaign. Second, to increase the impact of a campaign, the leader can also outsource a task to many elite Sybil workers, especially highly-ranked Dianping users. Finally, the leader actively participates in the review generation by directly generating the high-quality reviews by himself/herself or by closely supervising the review generation of workers. In summary, if elite Sybil workers are the puppets, then Sybil leaders are the masters who locally dominate the unique workflow of Sybil organizations on Dianping.

\subsubsection{Typical Workflow}\label{workflow}
\begin{figure}[t]
  \begin{center}
  \includegraphics[width = 0.45\textwidth]{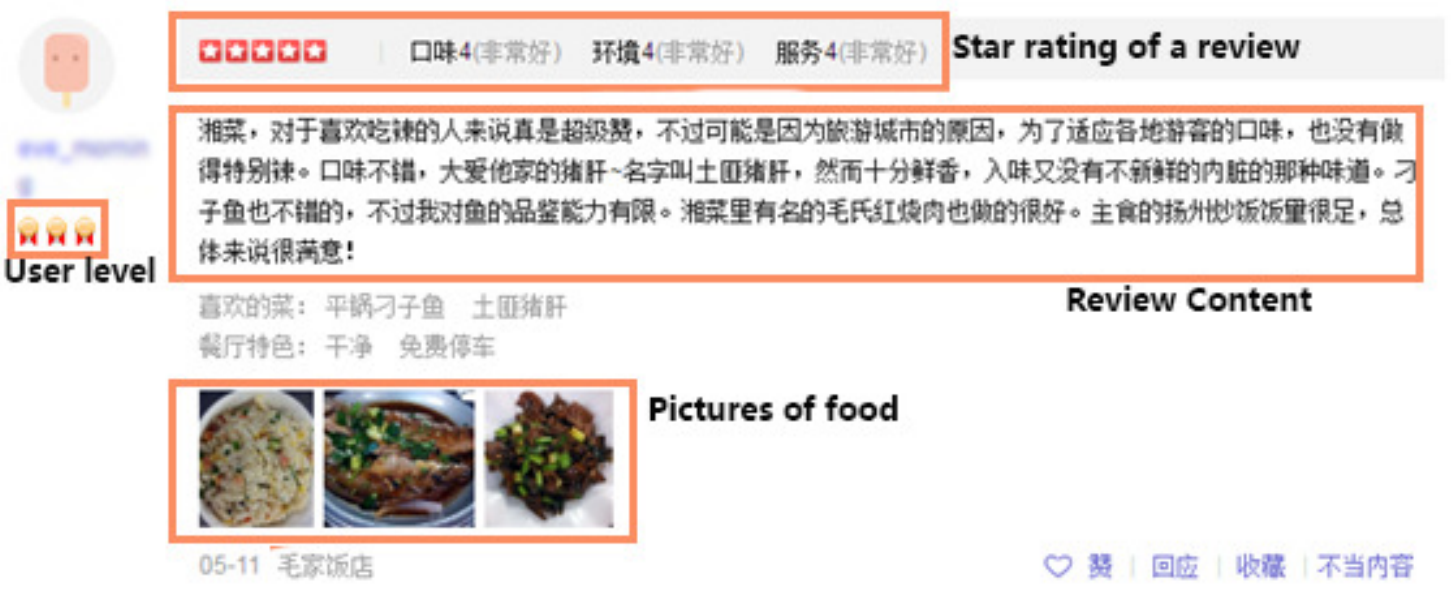}
  \caption{An example of a fake review}
  \label{fig:example}
  \end{center}
    \vspace{-0.6cm}
\end{figure}

Each Sybil campaign is centered on a collection of \textit{tasks}. For example, a campaign launched by an organization entails crafting positive fake reviews for a restaurant to boost ratings on Dianping. In this case, the owner of the overhyped store sets up the objects of a Sybil campaign, and the task is further distributed from organizers to Sybils. Each task would be ``posting a single (fake) positive review online.'' Sybils who complete a task generate \textit{submissions} that include screenshots of the fake reviews to be posted as evidence of his/her work (see Figure~\ref{fig:example}). The overhyped stores/agents can then verify if the work has been done to their satisfaction. It is important to notice that not all tasks can be completed because of some low-quality submissions.

\begin{figure}[htb]
  \begin{center}
  \includegraphics[width = 0.45\textwidth]{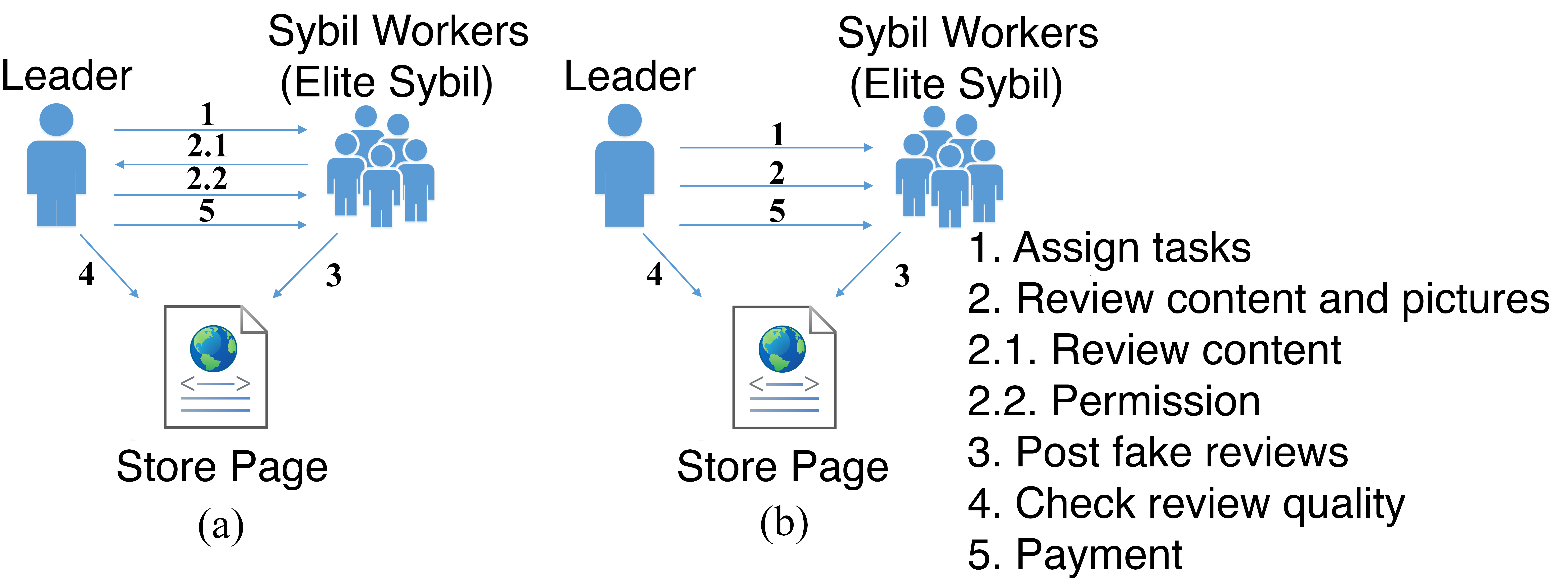}
  \caption{The process to post fake reviews}
  \label{Process of spam}
  \vspace{-0.2cm}
  \end{center}
\end{figure}

The key feature of a Sybil organization on Dianping is that the Sybil leader is actively involved in the Sybil tasks. In particular, when receiving a task from the customer, a Sybil leader distributes this task to multiple elite Sybil workers and guides review generation, which is illustrated in step (1) in Figure~\ref{Process of spam}.
\begin{itemize}
\item{\emph{Leader-supervised model}:} In this model, the reviews are created by an elite Sybil worker (step 2.1) and the generated content and posting time must follow the leader's guidance and must be approved by the said leader (step 2.2).

\item{\emph{Leader hands-on model}:} In this model, it is the leader or the customer that generates the review comments first. The generated reviews are normally high-quality comments that include both favored comments and pictures of food or the store (step 2).
\end{itemize}
Given a certain review, the worker posts fake reviews of the specified stores (step 3). The leader will check if these crafted fake reviews exist for a period of 3 to 7 days (step 4). Once the existence of fake reviews is confirmed, the leader will pay the elite Sybil worker (step 5).

Through our investigation, we find that cultivating a 3-star elite Sybil account endorsing a tutorial offered by a Sybil organizer is priced at \$6 per account. The tutorial provides details about the approach to boosting ratings of Sybil accounts and mimicking benign accounts. In concrete, (i) once an account is activated, its profile information, such as gender, date of birth, address, and profile picture, needs editing to the requirements of the tutorial. (ii) Before participating in the Sybil organization, a great number of reasonable reviews are also required to cultivate an elite Sybil account. Specially, through our sting operation in a Sybil organization on Taobao,\footnote{\url{https://www.taobao.com/}} the largest C2C website in China, we find that an overhyped store, through several Sybil organizers, have collected approximately $100,000$ elite Sybil accounts. For the Sybil organization we have infiltrated, we observed $30$ tasks which were assigned in three months. For a particular task, the store exploits some of these elite Sybil accounts to generate $500$ fake positive reviews at the cost of around \$3,000 in total. Moreover, the Sybil organization we participated in also provides an after-sales guarantee, meaning if fake reviews are deleted, it will launch a second-round elite Sybil attack. In addition, we also observe that rewards per submission on Dianping are many more than those on other Chinese websites, such as ZBJ and SDH~\cite{wang2012serf}. The high monetary rewards incentivize the Sybil agents or leaders to develop sophisticated pyramid schemes to evade detection. 

Based on the above discussion, it is clear that automatic detection of elite Sybil users is important to prevent Sybil attacks  from URSNs. This motivated us to develop a novel framework of Sybil detection.

\section{\textsc{ElsieDet}: Design and Implementation}\label{design_implementation}

\begin{figure}
  \begin{center}
  \includegraphics[width = 0.45\textwidth]{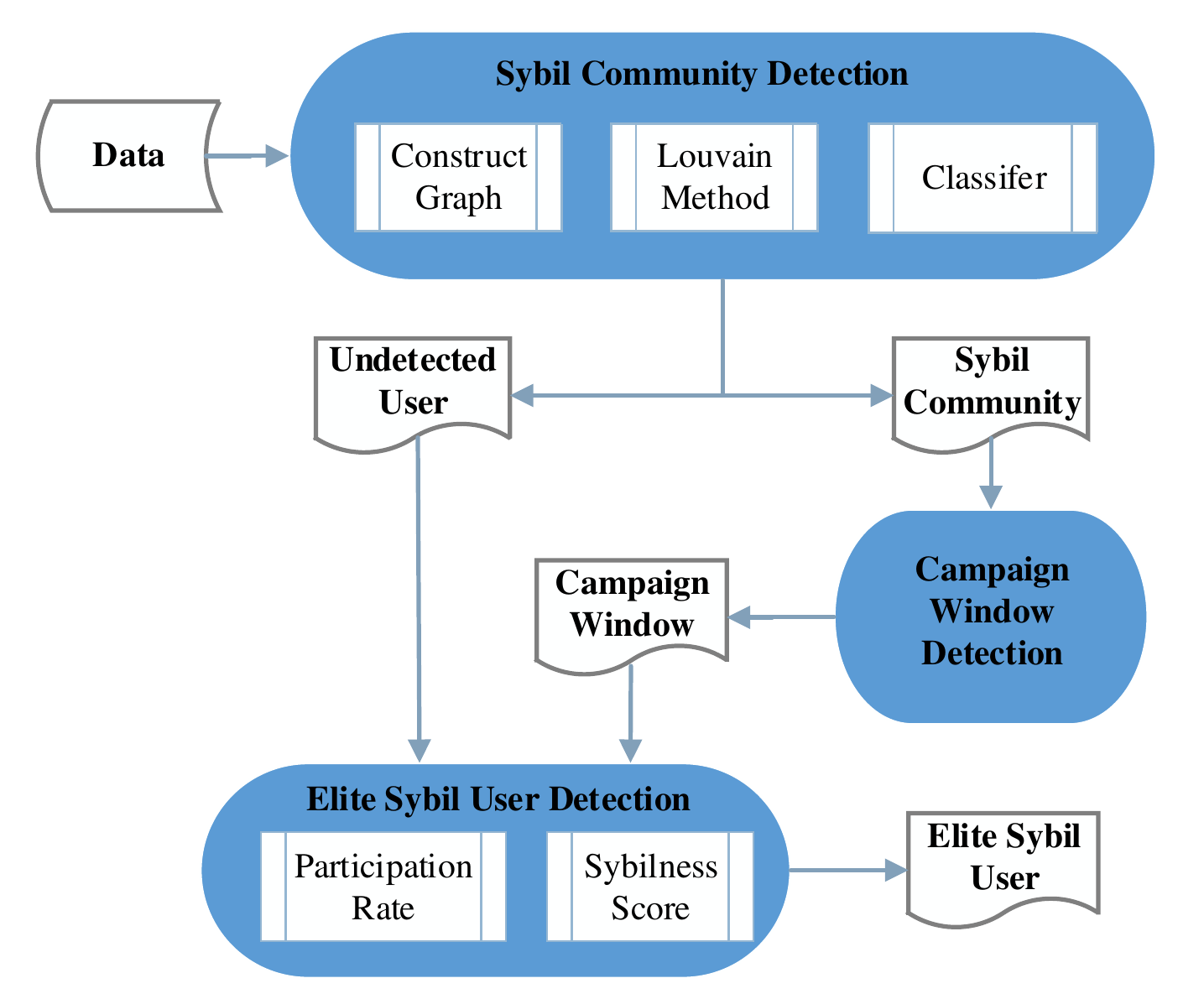}
  \caption{System overview}
  \label{fig:system framework}
  \vspace{-0.5cm}
  \end{center}
\end{figure}

In this section, we will present three components of \textsc{ElsieDet} (see Figure~\ref{fig:system framework}): detecting Sybil communities, determining campaign time windows, and detecting elite Sybil users. 

\subsection{System Overview}

\textsc{ElsieDet} is a three-tier Sybil detection system. In Phase~I, we cluster communities based on collusion networks and perform a binary classification on detected-communities, echoing that a large number of fake reviews are usually posted by the Sybil community under the guidance of the Sybil leaders.  
In this phase, regular Sybil users will be clustered in Sybil communities, but most elite Sybil users are able to evade community clustering by covering up their collusion.

In Phase~II, we extract time windows of Sybil campaigns from labeled Sybil communities. The rational behind the design is that a Sybil campaign has an active time period. A user posting a review towards the target store during the active time period is considered as a campaign-involved user. This user could be either a benign user who happens to visit the store and posts her reviews at that time or a Sybil user who posts fake reviews for the campaign benefits. We observe that a benign user posts reviews based on her real experience while a Sybil user always posts reviews during the active time period of the Sybil campaigns. 

In Phase~III, followed by the undetected users and corresponding extracted Sybil campaign time windows, we first use the participation rate between users and communities to characterize the extent to which a user is related to a community; then we leverage a novel metric to determine elite Sybil users. The rationale behind the elite Sybil detection algorithm is that, through using elaborate reviews to obfuscate their fake reviews, the elite Sybil users are motivated to participate in multiple Sybil campaigns due to a high economic rewarding. Therefore, the more campaigns a user participates in, the more likely the user is an elite Sybil user.

\subsection{Sybil Community Detection}\label{Organization_Detection}
In this section, we present the details of detecting the Sybil community. The first step of Sybil community detection is \emph{constructing the Sybil social links} for the Sybil users belonging to the same Sybil community. It is defined that two users belonging to the same community have a Sybil social link if they have \emph{collusive reviews}, which are similar reviews posted by two users according to the same store. Based on these virtual links, we further define a novel metric, \emph{pairwise similarity metrics}, which measure the similarity among the users.
Then, we adopt the Louvain method~\cite{Louvain2008fast} to extract communities from the entire network. Finally, we perform classification to identify the Sybil community from the benign community.

\subsubsection{\textbf{Constructing Sybil Social Links via Collusive Reviews}}
To cluster and identify the Sybil community, the first step is to build the social links between the Sybil users, which are coined as \textit{Sybil social links}. In general, two users belonging to the same community and having the \emph{collusive reviews} posted in the same store or restaurant are defined to have a Sybil social link. Specifically, 
a tuple abstraction of a user's single review is referred to as $(U, T, S, L)$, where $U$, $T$, $S$, and $L$ represent user ID, review timestamp, store ID, and star-rating of a review, respectively. For users $u$ and $v$, we derive \textbf{review sets} associated with $u$ and $v$, respectively:
\begin{equation*}\small
\begin{aligned}
\mathcal{R}(u) &= \{(U, T_1, S_1, L_1), (U, T_2, S_2, L_2), \cdots, (U, T_n, S_n, L_n)\};\\
\mathcal{R}(v) &= \{(V, T_{1}^{\prime}, S_{1}^{\prime}, L_{1}^{\prime}), (V, T_{2}^{\prime}, S_{2}^{\prime}, L_{2}^{\prime}), \cdots, (V, T_{m}^{\prime}, S_{m}^{\prime}, L_{m}^{\prime})\}.
\end{aligned}
\end{equation*}
For all pairwise users $u$ and $v$, and for a given $k$, $(U, T_k, S_k, L_k)\in \mathcal{R}(u)$, we define $P_{u}(k)=1$ if there exists $(V, T_{l}^{\prime}, S_{l}^{\prime}, L_{l}^{\prime})\in \mathcal{R}(v)$ such that the following three properties are true:
\begin{enumerate}\label{preperties}
\item The two reviews are posted in the same store: $S_k=S_{l}^{\prime}$;
\item The two reviews are created within a fixed time slot $\Delta T$: $|T_k - T_{l}^{\prime}| \leq \Delta T$;
\item Both two reviews are 1-star or both of them are 5-star: ${L_k} = L_l^{\prime} = \text{1-star}$ or ${L_k} = L_l^{\prime} = \text{5-star}$.
\end{enumerate}
Otherwise, $P_{u}(k)=0$.

Note that in previous research~\cite{CCS14syn}, Cao \emph{et al.} simply defined two collusive reviews if they pertain to the same constraint object and their timestamps fall into the same fixed time slot, but these two collusive reviews defined are not mathematically equivalent.

Measuring similarity is key to grouping similar users. Different from the previous research~\cite{CCS14syn, CCS14twitter} using Jaccard similarity metric, we measure the similarity between pairwise users $u$ and $v$ as follows:
\begin{equation}\small\label{similarity}
\begin{aligned}
\text{Sim}(u,v) &= \frac{\sum \limits_{k=1}^{n} P_{u}(k) + \sum \limits_{l=1}^{m} P_{v}(l)}{|\mathcal{R}(u)| + |\mathcal{R}(v)|}\\
&= \frac{\sum \limits_{k=1}^{n} P_{u}(k) + \sum \limits_{l=1}^{m} P_{v}(l)} {n + m}.
\end{aligned}
\end{equation}
Note: $\text{Sim}(u, v)=\text{Sim}(v, u).$

In summary, we model an Sybil community as an undirected weighted graph $\mathcal{G} = (\mathcal{V}, \mathcal{E})$, where each node $u \in \mathcal{V}$ is an user account and each edge $(u,v) \in \mathcal{E}$ represents a Sybil social link among users $u$ and $v$ if and only if $\text{Sim}(u, v) > \beta_\text{Thre}$.\footnote{The threshold $\beta_\text{Thre}$ is tuned to optimize the following community classification in terms of accuracy. Community classification results obtained by multiple supervised learning techniques are not overly-sensitive to the different thresholds chosen.} Then users $u$ and $v$ are defined as \emph{neighbors}.

\subsubsection{\textbf{Community Clustering via the Louvain Method}}
We then employ a community detection method, termed Louvain method~\cite{Louvain2008fast}, to detect communities on Sybil social links. The Louvain community detection method iteratively groups closely-connected communities together to improve the partition modularity. In each iteration, every node represents a community, and well-connected neighbor nodes are combined into the same community. The graph is reconstructed at the end of each iteration by converting the resulting communities to nodes and adding links that are weighted by the inter-community connectivity. The entire process is repeated iteratively until it reaches the maximum modularity. Each iteration has a computational cost linear to the number of edges in the corresponding graph and typically the process just requires a small number of iterations.

It is noted that community detection algorithms have been proposed to directly detect Sybils~\cite{viswanath2010analysis}. They seek a partition of a graph that has dense intra-community connectivity and weak inter-community connectivity. For example, the Louvain method searches for a partition with high modularity~\cite{Louvain2008fast}. However, we find that it is insufficient to uncover massive Sybil users within Louvain-detected communities. In the following step, we apply supervised machine learning to Louvain-detected communities.

\subsubsection{\textbf{Sybil Community Classification}}
Next, we apply machine learning classifiers to discriminate Sybil communities from benign ones.  The reason behind this is that some communities contain users who reside close-together or visit the same venues. To accurately characterize these observations, we apply eight features with respect to three types (tabulated in Table~\ref{Table:FeaturesList}) to our binary classifiers. The output is each community labeled either benign or Sybil. We validate this intermediate step in Section~\ref{results}.

\begin{table}[!htb]
\scriptsize
\caption{Types of features}
 \label{Table:FeaturesList}
 \centering
  \begin{tabular}{|c|c|}
 \hline {\bf  Types of Features}  & {\bf Features}  \\
 \hline \emph{Community-based Features}	  	           & \makecell{Score deviation, Average number of reviews, \\Entropy of the number of reviews in each chain \\stores, Entropy of districts of stores}\\
  \hline \emph{Network Features}             & \makecell{Average similarity, \\Global clustering coefficient} \\
  \hline \emph{User-based Features} 	          & \makecell{Unique reviews ratio, \\Maximum number of duplication} \\
  \hline
 \end{tabular}
\end{table}

\noindent \textbf{(a) Community-based features.} There are four types of Community-based features: score deviation, reviews per store, entropy of chain stores, and entropy of districts of stores.
\emph{Score deviation} and \emph{Average number of reviews} are self-explanatory. To achieve the Sybil tasks, score deviation of reviews posted by Sybil users will become larger.
\emph{Entropy of the number of reviews in each chain stores} is the expected value of information contained in each of the chain stores by measuring the number of reviews occurred. We use this feature because some Sybil users post reviews only in chain stores.
\emph{Entropy of districts of stores} is a location-based feature to characterize mobility patterns of Sybil users that are driven by Sybil tasks. We therefore use \emph{Entropy of districts of stores} to show this difference.

\begin{figure}[h!]
  \centering
  \mbox{
    \subfigure[Graph structure of a benign community (community $7914$)\label{fig:benign community 7914}]{\includegraphics[width=0.22\textwidth]{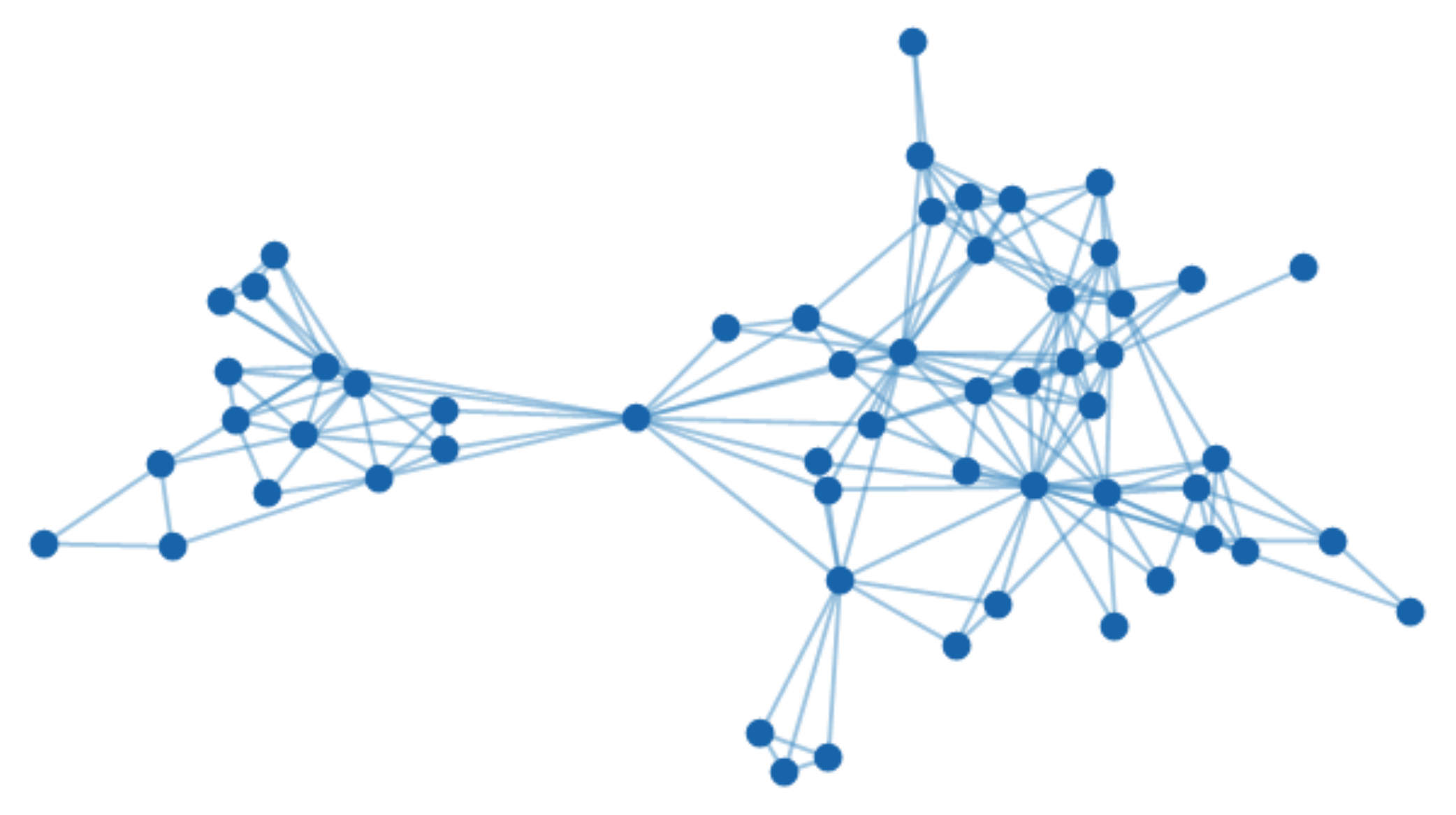}}\quad
    \subfigure[Graph structure of a Sybil community $7924$ (community $7924$)\label{fig:sybil community 7924}]{\includegraphics[width=0.22\textwidth]{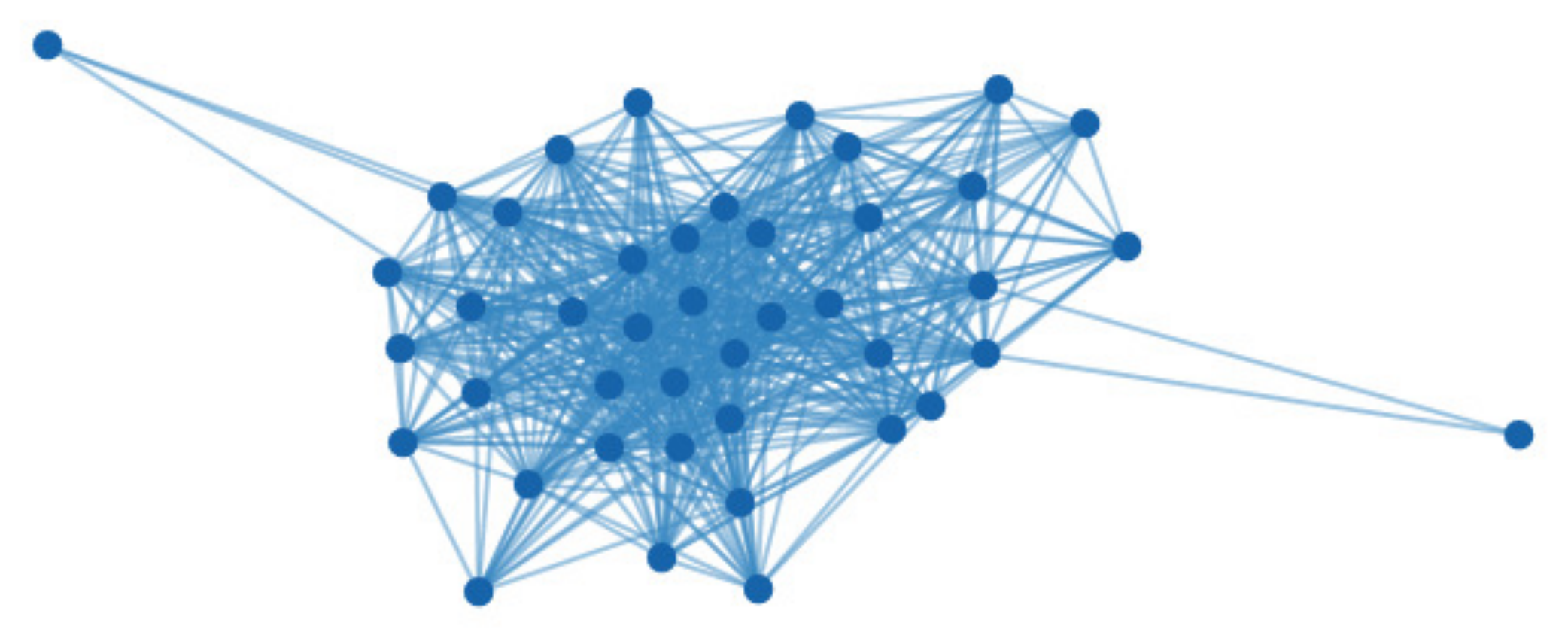}}
    }
  \caption{Comparison of the graph structure between a benign community and a Sybil community}
  \label{fig:sybil community}
\end{figure}

\noindent \textbf{(b) Similarity-based network features.} We redefine the network via Sybil social community construction since benign and Sybil communities have remarkable differences with respect to the graph structure (see Figure~\ref{fig:benign community 7914} and Figure~\ref{fig:sybil community 7924}). 
We use \emph{Average similarity} and \emph{Global clustering coefficient} to show the difference according to the redefined graph structures.
\emph{Average similarity} is the average similarity between pairwise users in a community. Sybil users in a Sybil community are assigned tasks for similar stores, but users in a benign community randomly choose stores to post reviews. Hence, similarity values between Sybil users are greater than those between benign users.
\emph{Global clustering coefficient} is used to measure the degree in which nodes in a graph tend to cluster together. Sybil users have the characteristics of team working, so they are more likely to be clustered together.

\noindent \textbf{(c) User-based features.} Since community-based features may lose information of users, we then abstract the user-based features of each user and aggregate them as a feature of the community. By analyzing Sybil communities, we observe that some Sybil users will repeatedly post reviews in the same store. We therefore define two features, \emph{Unique reviews ratio} and \emph{Maximum number of duplication}, to reflect this user-level behavior. Lastly, we do not use linguistic or contextual features because these features are not so effective in the URSN setting~\cite{mukherjee2013yelp}.

\subsection{Campaign Window Detection}\label{Campaign_Detection}

To detect the time window of a Sybil campaign, one potential approach is detecting sudden increases in rating, ratio of singleton reviews, and the number of reviews by leveraging a burst detection algorithm (\emph{e.g.}, Bayesian change point detection algorithm~\cite{erdman2007bcp}). However, on Dianping, Sybil campaign detection results based on burst detection may not be reliable in practice. For example, the sudden increases in ratings or the number of reviews may be contributed by some unexpected factors such as offline promotions. An observation is that a store tends to entice its customers to write favorable reviews as the return of a discount coupon in promotion seasons. 

Different from the previous research, our proposed solution focuses on detecting the anomaly collaborative behaviors of Sybil community. We interpret the algorithm of campaign window detection in the following. 
The Algorithm~\ref{alg:Campaign Window Detection} takes as input a list $L_{\mathit{review}}$ that represents the number of reviews posted each week and does the following:
\begin{enumerate}
\item Initializes the start and end points of the campaign window (Line 1 through Line 2).
\item Iteratively finds and deletes sparse review intervals within the campaign window (Line 3 through Line~14).
        \begin{enumerate}
        \item Finds the first left and right sparse review intervals within the campaign window. If none, the functions will return the entire campaign window (Line 4 through Line 5).
        \item If there is no sparse review interval on either side, breaks the loop (Line 6 through Line~8).
        \item Removes the sparse review interval. This can prevent deleting major parts of the campaign window (Line 9 through Line 13).
        \end{enumerate}
\end{enumerate}
The output of Algorithm~\ref{alg:Campaign Window Detection} is the start point and the end point of each Sybil campaign accordingly.



\begin{algorithm}[h]
\scriptsize
\caption{\textbf{Detecting Campaign Time Windows}}\label{alg:Campaign Window Detection}
\begin{algorithmic}[1]
\renewcommand{\algorithmicrequire}{\textbf{Input:}}
\renewcommand{\algorithmicensure}{\textbf{Output:}}
\REQUIRE A list $L_{\mathit{review}}$ whose item $L_{\mathit{review}}[i]$ denotes the number of reviews posted in the $i$th week.
\ENSURE The start point $l$ and end point $r$ of the campaign time window.
\\ \textit{Initial}:
 \STATE $\mathit{l} \gets 0$;\\
 \STATE $\mathit{r} \gets \mathit{length}(L_{\mathit{review}}) - 1$;\\
 \WHILE{(\TRUE)}
   \STATE ${\mathit{I}}_{\mathit{l,l'}}  \gets \mathit{find}(\mathit{left}, l)$; \COMMENT{Find the first sparse interval $I_{l,l'}$ from left.}\\
  \STATE ${\mathit{I}}_{\mathit{r',r}} \gets \mathit{find}({\mathit{right}}, r)$; \COMMENT{Find the first sparse interval $I_{r',r}$ from right.}\\
	 \IF [There is no sparse interval.]{(${\mathit{l'}} = \mathit{r}$ \AND
   ${\mathit{r'}} = \mathit{l}$)}
		 \STATE \textbf{break};
     \ENDIF
	 \IF [Choose the interval with fewer reviews.]{($|{\mathit{I}}_{\mathit{l,l'}}| \le |{\mathit{I}}_{\mathit{r',r}}|$)}
		 \STATE $\mathit{l} \gets {\mathit{l'}} + 1$;\\
     \ELSE
     	 \STATE $\mathit{r} \gets {\mathit{r'}} - 1$;\\
	 \ENDIF
 \ENDWHILE
 \RETURN $\mathit{l}, \mathit{r}$;\\
\end{algorithmic}
\end{algorithm}

\begin{figure}[t]
  \begin{center}
  \includegraphics[width = 0.475\textwidth]{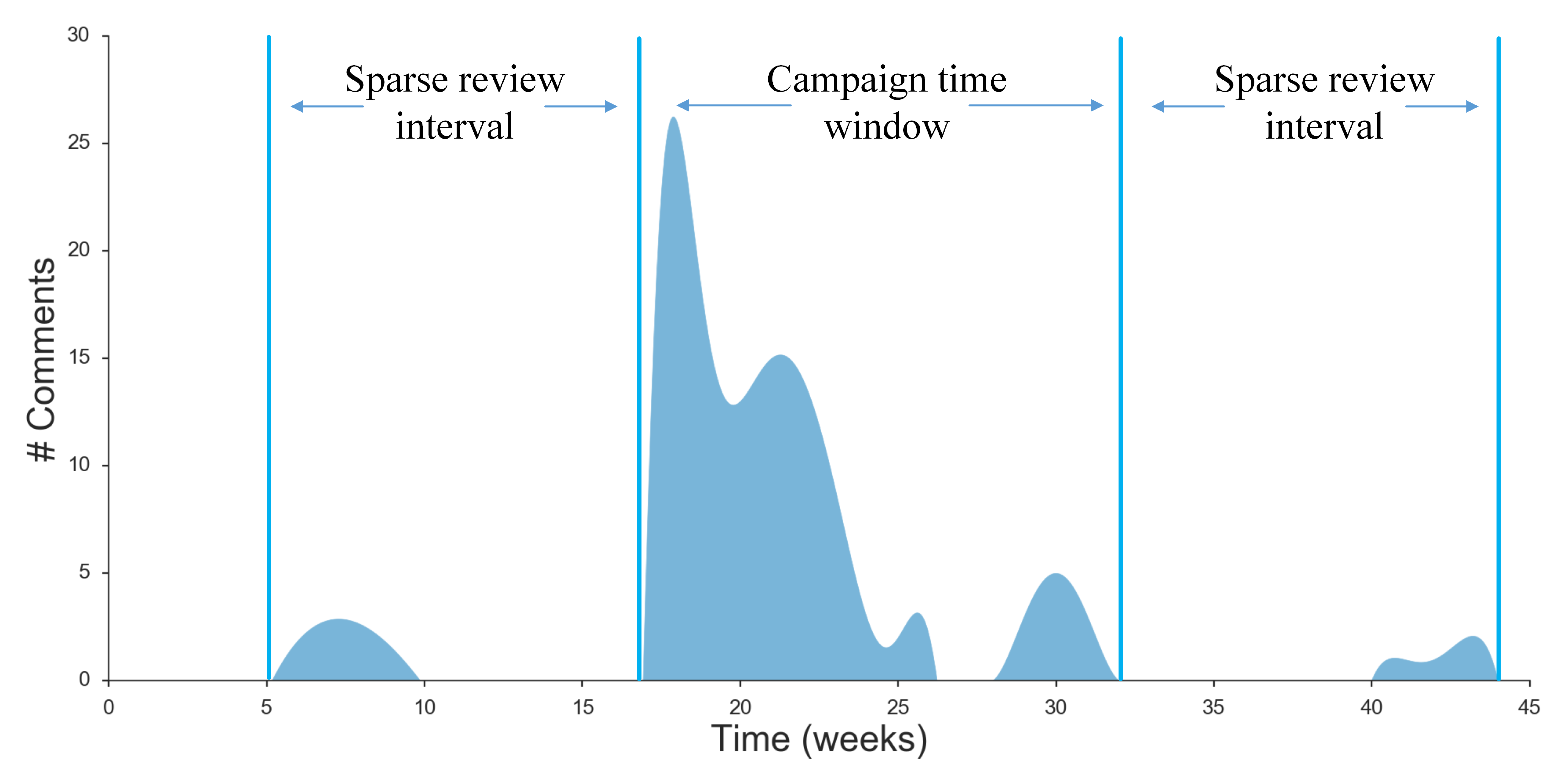}
  \vspace{-0.3cm}
  \caption{An example of campaign time window detection}
\label{fig:campagin time window detection}
\vspace{-0.5cm}
  \end{center}
\end{figure}

As shown in Figure~\ref{fig:campagin time window detection}, it is observed that a campaign period is comprised of multiple segment periods. We are interested in those segment periods in which the Sybil users are active and thus we need to filter out those periods when the Sybil users are inactive. To achieve this, we introduce the concept of Sparse Review Interval, which is used to indicate whether or not the users are active in this time period. In particular, \textit{a sparse review interval $I_{i,j}$ (where $i$ represents the start point of the $i$th week and $j$ represents the end point of the $j$th week) is referred to as the period in which the number of weeks with at least one review is less than the number of weeks without any reviews.} As shown in Figure~\ref{fig:campagin time window detection}, with a long time period, the entire time interval can be seen as a sparse review interval. In order to avoid removing intervals with massive reviews, our strategy is scanning the time period from both left and right to find the first sparse review intervals respectively, and then removing the sparse review interval with fewer reviews. We repeat this process until there is no sparse review interval and the remaining period is the targeted campaign period.

\subsection{Elite Sybil User Detection}\label{SenoirSybil_Detection}
Recall that elite Sybil users are those who often post reviews not belonging to Sybil tasks like a benign user but occasionally post fake reviews.
The primary reason that the existing Sybil detection approaches cannot effectively detect elite Sybil users is that reviews not belonging to Sybil tasks decrease the similarity between elite Sybil users. Labeling all reviews of an elite Sybil user as fake reviews may misjudge some real reviews, which will take away the enjoyment of the service. In order to detect elite Sybil users, we take as input the time windows of Sybil campaigns and corresponding undetected users. Then we define the participation rate and \textit{Sybilness} that is the perceived likelihood to output an elite Sybil user. Finally we use \textit{Sybilness} to quantify each review.

\noindent \textbf{Participation rate between users and communities.} We first define participation rate between users and communities to characterize the extent to which a user is related to a community. Based on our observations, we assume that the more campaigns a user participates in, the more likely the user is an elite Sybil user. Given a community $C$, we define:
\begin{itemize}
\item $N_{C}(k)$: the accumulated number of reviews posted within the $k$th time window of community $C$.
\item $N_{C}^{\max}$: the maximum number of reviews posted within all time windows of community $C$.
\end{itemize}
We then ``normalize'' the number of reviews in the $k$th time window by $P_{C}(k)=\frac{N_{C}(k)}{N_{C}^{\max}}$, for a given $C$. $P_{C}(k)$ will help indicate the importance of a time window, since the larger the number of reviews is within a time window, the more active this campaign is in the community. Then for a given user $u$ in community $C$, we can calculate the ``weighted sum'' of the number of reviews $u$ posts by:
\begin{equation}\small
  N_{u \in C}=\sum_k{P_{C}(k) \cdot N_{u \in C}(k)},
\end{equation}
where $N_{u \in C}(k)$ represents the number of reviews $u$ posted within the
$k$th time window of $C$. We finally plug $N_{u \in C}$ into a standard sigmoid function to measure the participation rate $\rho_{u \in C}$ between $u$ and $C$:
\begin{equation}\small
  \rho _{u \in C} = \frac{1}{1+\exp^{-\frac{N_{u \in C} - \mu_C}{\sigma_C}}}, \quad \text {for any}~u \in C,
\end{equation}
where $\mu_C$ and $\sigma_C$ are the mean and the variance of $N_{u \in C}$ for all users~$u$ in $C$.

\noindent \textbf{Sybilness.} \textit{Sybilness} score is a perceived likelihood indicating if a user is an elite Sybil user.
Since simultaneously participating in multiple communities leads to the large cardinality of $C$ but small $N_{u \in C}$ and $\rho _{u \in C}$, just considering about the participation rate $\rho _{u \in C}$ will fail to tease out elite Sybil users.
We then take $\rho _{u \in C}$ into consideration to construct the final index, \textit{Sybilness}, to determine a specific user's legitimacy. To be specific, for assigning a \textit{Sybilness} score $f$ to each user $u$ on Dianping, we take a weighted average method on $N_{u \in C}$ with respect to each of the corresponding coefficients $\rho _{u \in C}$, for all~$C$, as shown below:
\begin{equation}\small
\begin{aligned}
\label{score}
f(u)=\sum_{C}{\rho _{u \in C}\cdot N_{u \in C}}.
\end{aligned}
\end{equation}
Eventually, we use the \textit{Sybilness} score $f(\cdot)$ to determine the perceived likelihood that a user is an \textit{elite Sybil user} or not (Note: \textit{Sybilness} score here can be greater than $1$.).

\noindent \textbf{Annotating reviews posted by elite Sybil users.}
Since not all reviews posted by elite Sybil Users are fake, we annotate each reviews with a score defined as $\rho _{u \in C}\cdot{P_{C}(k)}$, for any $k$. This score can be used as a criterion to filter fake reviews or regulate the frequency of CAPTHCHAs.

\section{Evaluation}\label{section:evaluation}
We implement \textsc{ElsieDet} and evaluate it on a large-scale dataset of Dianping. Our evaluation covers the following aspects: Sybil community detection, elite Sybil user detection, and system performance.

\subsection{Data Collection}
In this section, we will introduce the datasets used and propose the methodology we use to gain the ground-truth data.

\begin{table}[h!]\footnotesize
\caption{Breakdowns of stores}
\label{Table:intro-distribution}
    \centering
     \begin{tabular}{|c|c|c|c|}
    \hline
    \textbf{Type} & \textbf{\# Stores} &  \makecell{\bf \# Overhyped \\ \bf Stores} & \makecell{\bf Percentage of \\ \bf Overhyped Stores}\\
    \hline
    Cinema & $\cnum{235}$  & $\cnum{71}$ & $\mathbf{30.21}\%$\\
    \hline
    Hotel & $\cnum{1738}$ & $\cnum{134}$ & $\pnum{7.71}$\\
    \hline
    Restaurant & $\cnum{22474}$ & $\cnum{1244}$ & $\pnum{5.54}$\\
    \hline
    Entertainment & $\cnum{1384}$ & $\cnum{73}$ & $\pnum{5.27}$\\
    \hline
    Wedding Service& $\cnum{320}$ & $\cnum{8}$ & $\pnum{2.50}$\\
    \hline
    Beauty Store & $\cnum{1460}$ & $\cnum{35}$ & $\pnum{2.40}$\\
    \hline
    Fitness Center & $\cnum{326}$ & $\cnum{7}$ & $\pnum{2.15}$\\
    \hline
    Living Service & $\cnum{863}$ & $\cnum{10}$ & $\pnum{1.16}$\\
    \hline
    Scenic Spots & $\cnum{1243}$ & $\cnum{14}$ & $\pnum{1.13}$\\
    \hline
    Shopping & $\cnum{2466}$ & $\cnum{22}$ & $\pnum{0.89}$\\
    \hline
     Infant Service & $216$ & $0$ & $\pnum{0}$\\
   \hline
    Car & $148$ & $0$ & $\pnum{0}$ \\
   \hline
    Decoration Company & $67$ & $0$ & $\pnum{0}$ \\
    \hline
    \end{tabular}
    \vspace{-0.1cm}
\end{table}

\noindent \textbf{Dataset.}
We develop a Python-based crawler to analyze HTML structure of store pages and user pages on Dianping. All reviews were crawled by the web crawler from January 1, 2014 to June 15, 2015.  Starting from the four hand-picked overhyped stores (the seed list) in the training set belonging to the same Sybil organization, which we discovered during our month-long investigation. We then crawled outwards---crawling one level down of all users who wrote reviews in these stores and extended the store list that was commented by these users. Second, we crawled all reviews appearing in these stores and collected all users of these reviews to form a user list. The web crawler repeated these steps until reaching $32,940$ stores on the store list. Eventually, our resulting data set has $10,541,931$ reviews, $32,933$ stores, and $3,555,154$ users. We will make all of our data used publicly available in the future. Furthermore, we categorize the stores crawled into $13$ types (see breakdowns in Table~\ref{Table:intro-distribution}). In Table~\ref{Table:intro-distribution}, the $13$ categories are shown in decreasing order in terms of percentage of overhyped stores. Followed by our detection methodology, surprisingly, we find that more than $30\%$ overhyped stores are pertinent to cinemas. The main remaining overhyped stores are hotels, restaurants, and places of entertainment.

\noindent \textbf{Ground-truth dataset.}
Similar to the previous research~\cite{CCS14twitter, www12fim, egele2013compa}, we rely on manually labeled data for Sybil community detection. In order to classify the communities as benign or Sybil using supervised learning, a subset of the communities needs to be labeled. To carry out the labeling, 
we actively exchanged ideas with Dianping of how high-profile Sybil users resemble. Particularly, the final manual labeling considers the following three criteria. 
If two of them are satisfied, then a community is labeled as a Sybil community.

\noindent \textbf{\emph{(a) Massive filtered reviews by Dianping}} signify that a large proportion of reviews posted in a community are filtered by Dianping's Sybil detection system. Reviews that Dianping has classified as illegitimate using a combination of algorithmic techniques, simple heuristics, and human expertise. Filtered reviews are not published on Dainping's store/user pages. If we find a great proportion of reviews existing in our dataset but missing on Dianping's main listings, this indicates that these reviews have been filtered. 
Although a review can be filtered for many reasons, such as overly-florid or low-quality reviews, filtered reviews are, of course, partial indicators of being Sybil. If massive reviews have been filtered in a community, then the community has a high possibility to be Sybil.

\noindent \textbf{\emph{(b) Duplicate user reviews}} mean that reviews posted by a user belonging to a community only serve one or two store(s) with similar content. To our observation, reviews posted by a benign user of a community are often evenly distributed in miscellaneous stores. The existence of \emph{duplication} signifies that Sybil users are more addicted to boosting review ratings in only a few stores in a community. This feature is stricter than the collusive reviews defined in this paper.

\noindent \textbf{\emph{(c) Spatio-temporal review pattern}} means that an unusual sudden jump with respect to the number of reviews of a target restaurant/store in a community is consistent with a collusive action of the Sybil community, by rule of thumb. Normally, the reviews of a store are evenly distributed since its inception. 
Hence, if many stores appearing in a community demonstrate unreasonable spatio-temporal patterns, then the community is highly likely to be Sybil.

To do this, we did not hire Amazon Mechanical Turk (AMT) to accomplish the tasks because scrutinizing those reviews requires deep familiarity with Chinese language and the Dianping platform~\textit{per se}. Instead, we hired $5$ Chinese undergraduate students to classify communities as either benign or Sybil. For the rare cases where there was not a consensus, we used voting. For example, a community would be labeled as Sybil if and only if the $5$ votes are SSSBB, SSSSB or SSSSS, with S representing Sybil and B representing benign.



\subsection{Results and Detection Accuracy}\label{results}

\noindent \textbf{Accuracy of Sybil community detection.}
For the dataset used, \textsc{ElsieDet} detects in total $710$ communities. 
By using the multiple criteria shown above, we randomly picked up $170$ communities as ground truth and labeled $117$ Sybil communities as well as $53$ benign communities. The assumption that a community only takes a binary classification can be justified by the empirical percentage of Sybil (resp. benign) users taking up in the designated Sybil (resp. benign) communities. To justify this, we took a look at each of the $1,969$ users of $74$ communities ($54$ Sybil vs. $20$ benign) obtained from ground truth (which is more than $10\%$ of the total amount of communities), still by following the above criteria to check each user in communities. We conclude that $ 96.85\%$ of the users are designated to the correct community labels. With $8$ features tabulated in Table~\ref{Table:FeaturesList}, we also compare several classifiers implemented by \emph{scikit-learn} library~\cite{sklearn}. We perform grid search to determine optimal parameters for each classifier and evaluate their performance on weighted \emph{precision}, weighted \emph{recall}, weighted \emph{F1 score}, and \emph{AUC} (Area under the Curve of ROC) using 5-fold cross-validation. As shown in Table~\ref{Table:Classifiers}, support vector machine (SVM) performs best among all classifiers with $96.45\%$ F1 score and $99.42\%$ AUC, using Gaussian (RBF) kernel with parameters chosen $\mathtt{C} = 18$ and $\mathtt{\gamma} = 0.09$.

\begin{table}[!htb]\small
\vspace{-0.1cm}
 \caption{Classification performance}
 \label{Table:Classifiers}
 \centering
 \begin{tabular}{|c|c|c|c|c|}
 \hline
 {\bf  Classifier}  & {\bf Precision} & {\bf Recall} & {\bf F1} & {\bf AUC} \\
 \hline
 \hline
 \emph{Decision tree} & $\pnum{93.80}$ & $\pnum{92.90}$ & $\pnum{93.60}$ & $\pnum{92.83}$ \\
 \hline
  \textbf{SVM} & $\pnum{96.74}$ & $\pnum{96.47}$ & $\mathbf{96.45}\%$ & $\mathbf{99.42}\%$ \\
 \hline
  \emph{GNB} & $\pnum{94.21}$ & $\pnum{93.44}$ & $\pnum{93.57}$ & $\pnum{97.64}$ \\
 \hline
  \emph{KNN} & $\pnum{96.75}$ & $\pnum{96.47}$ & $\pnum{96.50}$ & $\pnum{97.45}$ \\
 \hline
  \emph{Ada boost} & $\pnum{93.84}$ & $\pnum{93.54}$ & $\pnum{93.60}$ & $\pnum{97.92}$ \\
 \hline
  \emph{Random forest} & $\pnum{93.16}$ & $\pnum{94.01}$ & $\pnum{92.99}$ & $\pnum{97.42}$ \\
 \hline
 \end{tabular}
\end{table}

We then apply our trained classifiers to predict each community. As a result, \textsc{ElsieDet} identifies $566$ Sybil communities with $22,324$ users, and $144$ benign communities with $5,222$ users. Surprisingly, detected Sybil communities significantly outnumber detected benign communities. It is perhaps because in the community clustering process, the constraints of posting time and review ratings pose limitations on forming benign communities. Most benign users are thereby pruned by applying the Louvain method.
\begin{figure}[t]
  \begin{center}
  \includegraphics[width = 0.35\textwidth]{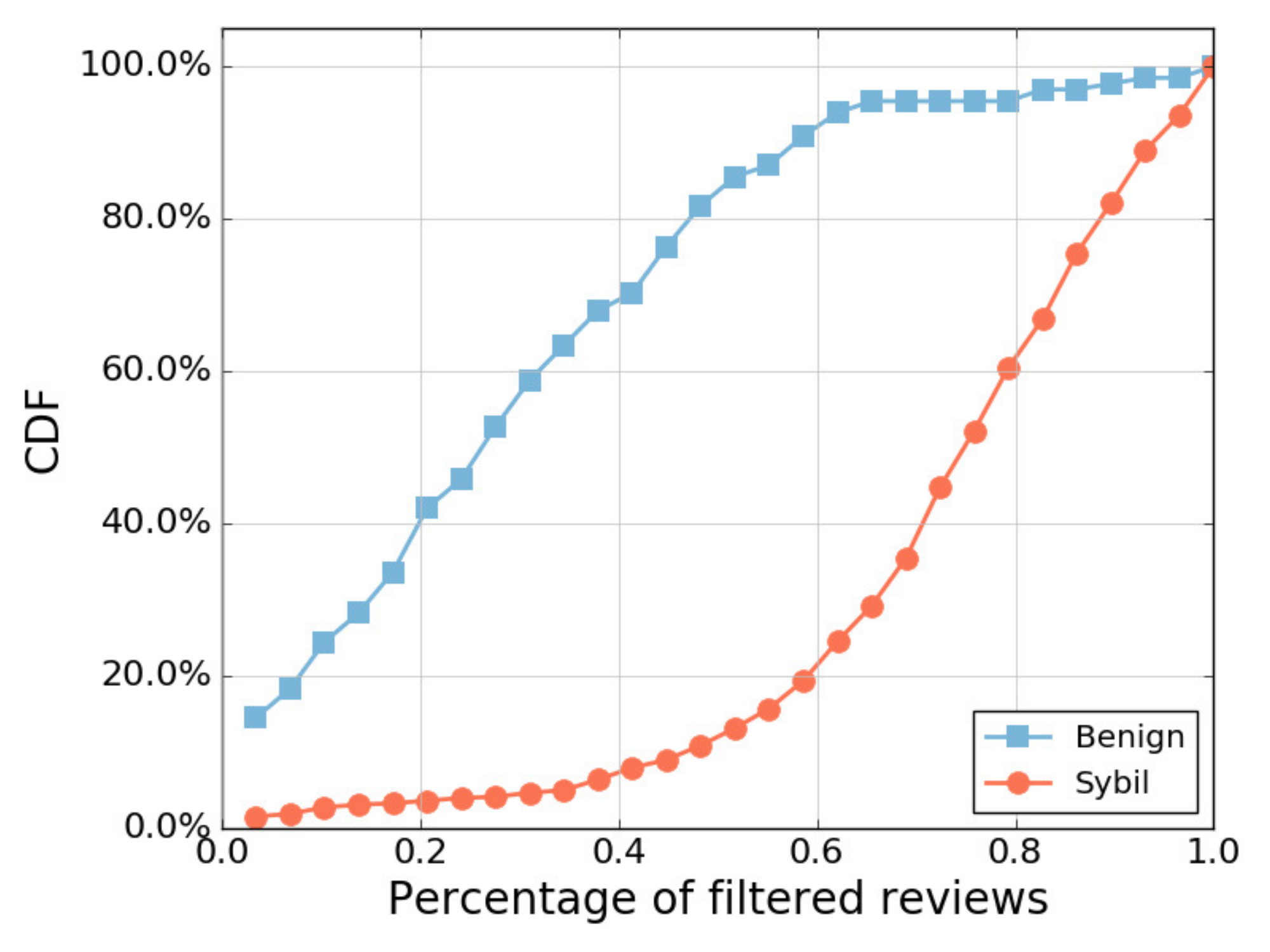}
  \vspace{-0.3cm}
  \caption{Comparison between benign and Sybil communities by percentage of filtered reviews}
  \label{fig:CDF-filtered-reviews-seed}
  \vspace{-0.4cm}
  \end{center}
\end{figure}

\begin{figure*}[htb]
  \centering
  \mbox{
    \subfigure[Comparison on the number of fake reviews\label{fig:comparison fake reviews}]{\includegraphics[scale = 0.28]{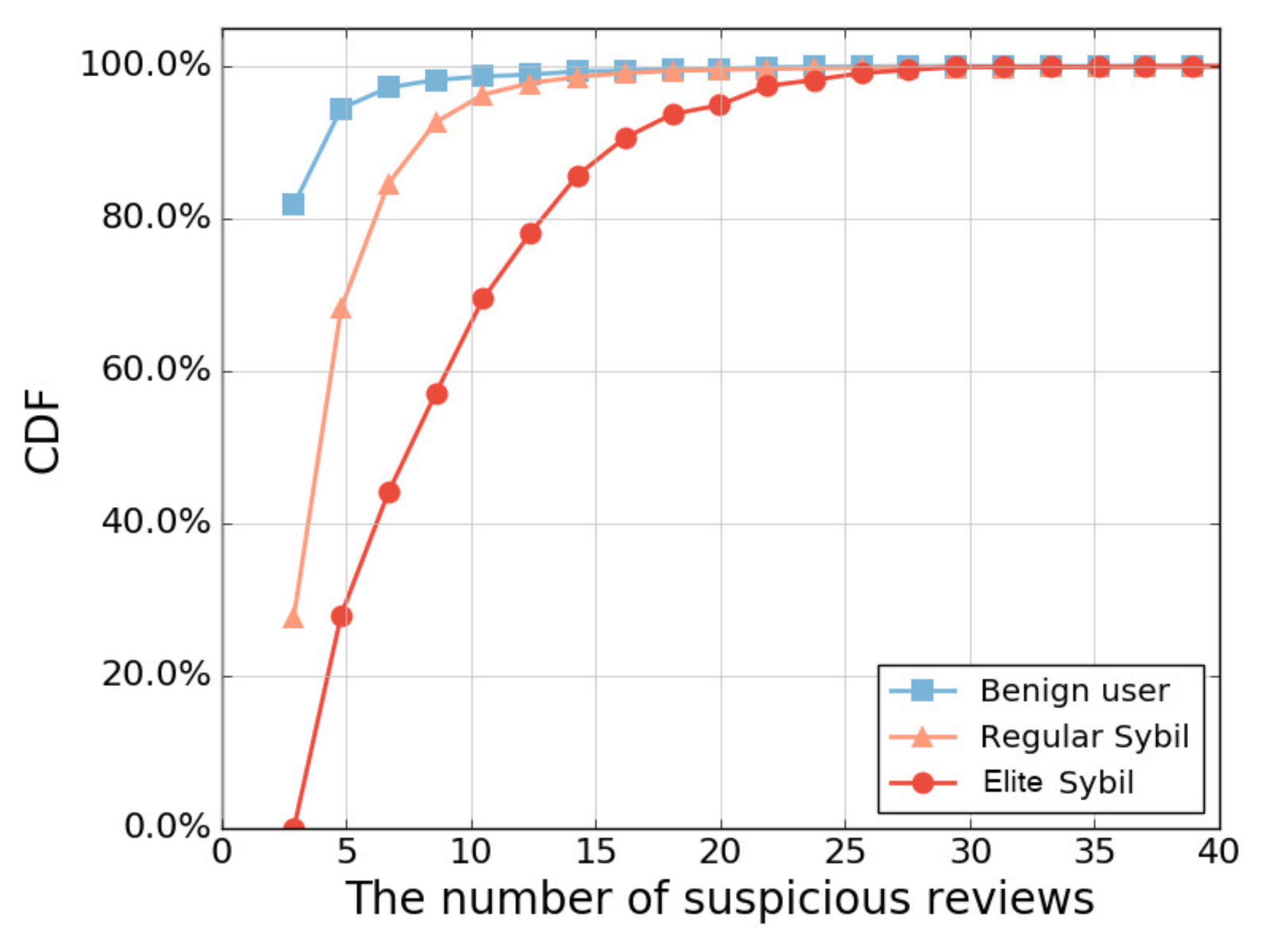}}
    \subfigure[Comparison on the percentage of fake reviews\label{fig:percentage fake reviews}]{\includegraphics[scale = 0.28]{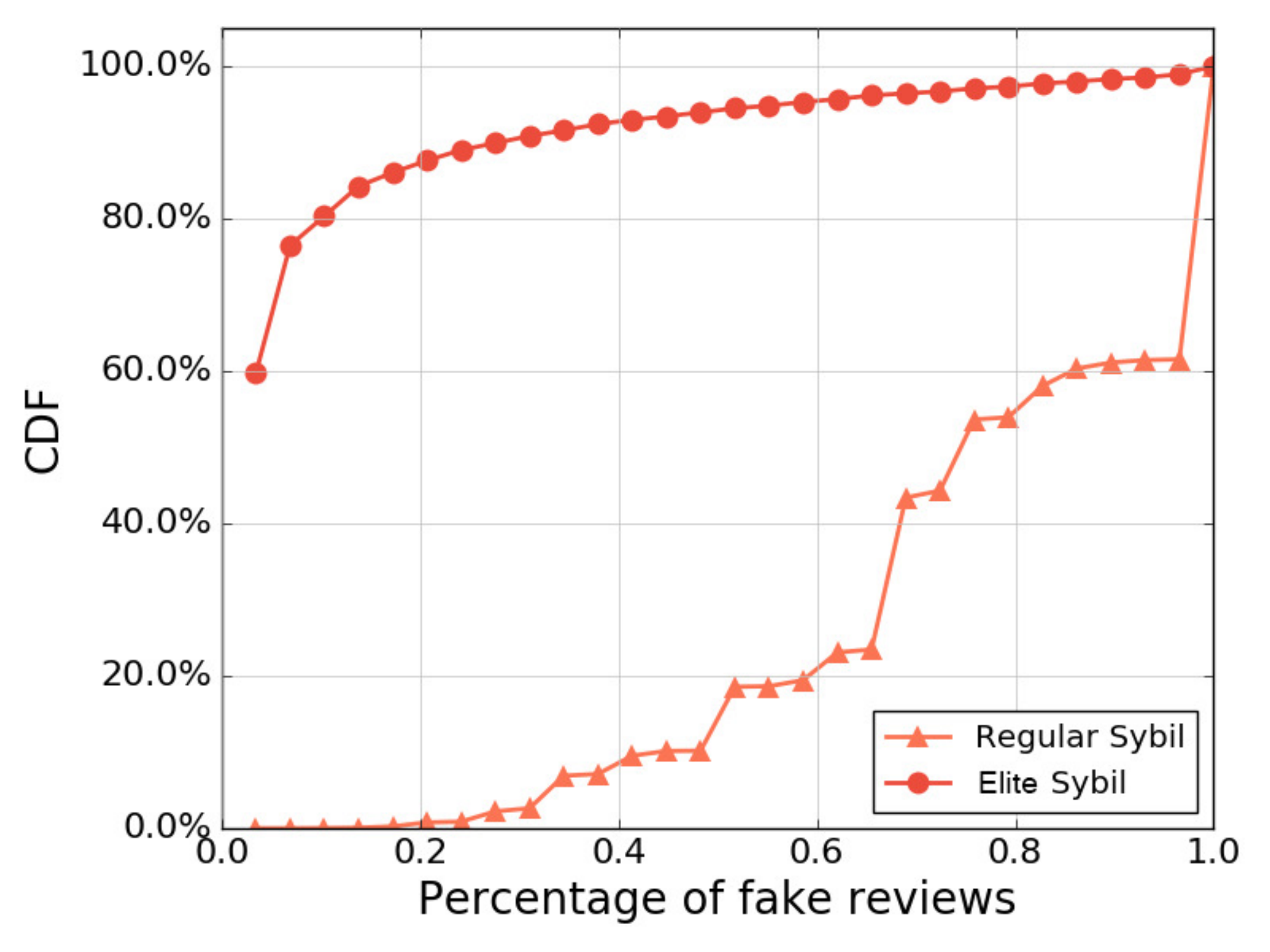}}
    \subfigure[Comparison on percentage of filtered reviews\label{fig:comparison filtered reviews}]{\includegraphics[scale = 0.28]{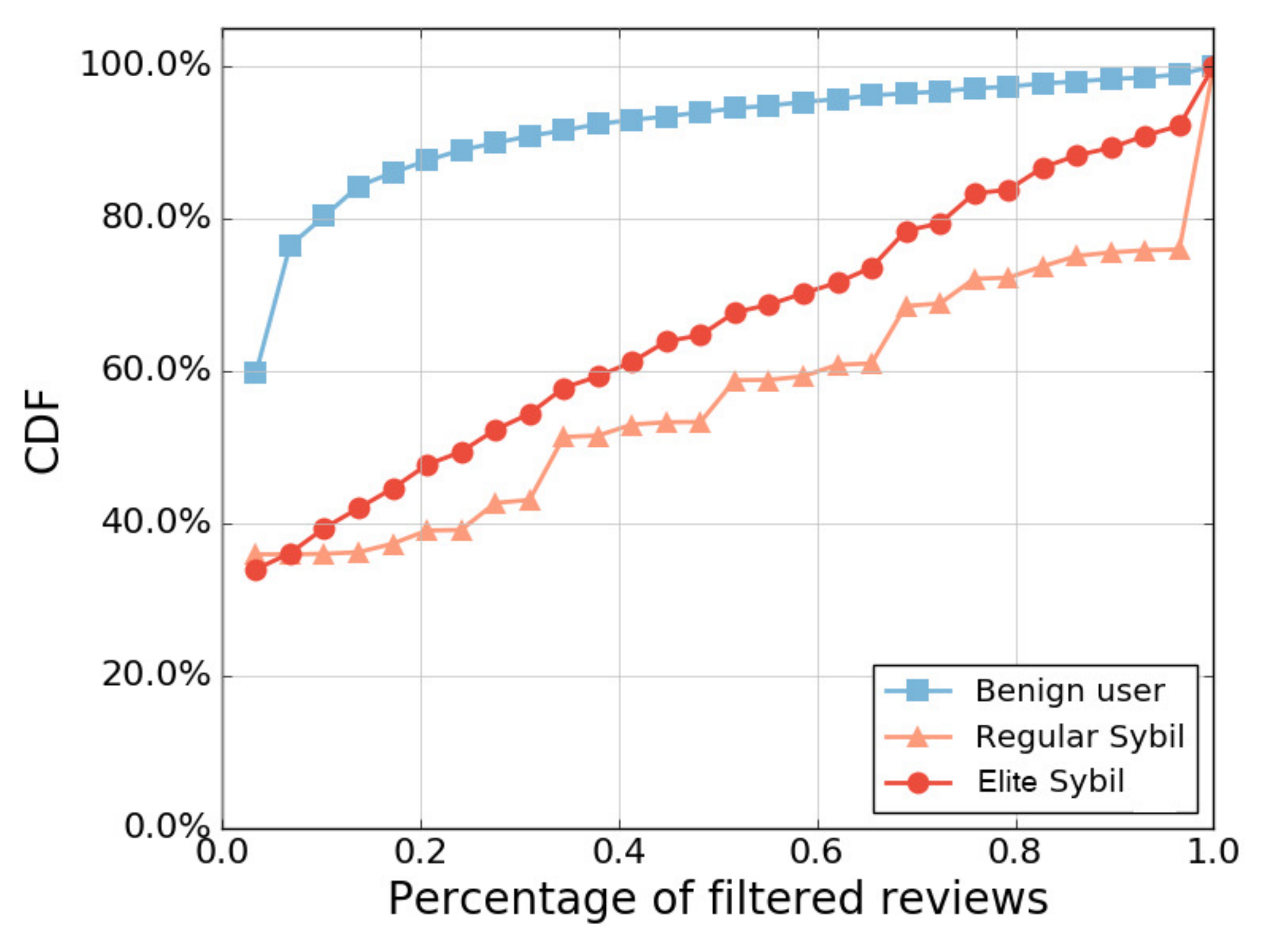}}
    }
    \vspace{-0.3cm}
  \caption{Comparison between elite Sybil users and regular Sybil users}
  \label{fig:Comparison senior Sybil regular Sybil}
  \vspace{-0.2cm}
\end{figure*}

Recall in Section~\ref{ssec:Dianping}, we note that not all filtered reviews are fake reviews (some are viewed as useless reviews). Through our experiments, we confirm that the fake reviews classified are more likely to be filtered. As shown in Figure~\ref{fig:CDF-filtered-reviews-seed}, we compare the percentage of filtered reviews in benign and Sybil communities, respectively. We observe that the percentage of filtered reviews of Sybil communities significantly outweighs that of benign communities with respect to the same cumulative probability. Specifically, we see that $80\%$ of Sybil communities (resp. benign communities) have more than $80\%$ (resp. less than $50\%$) of reviews filtered. We conclude that filtered reviews are more likely fake, which validates the accuracy of our detection methodology.

\noindent \textbf{Accuracy of elite Sybil users detection.}
\textsc{ElsieDet} considers a user~$u$ as an elite Sybil user if the following two conditions hold: (i) if the user~$u$ does not belong to any community; and (ii) the user participation rate $\rho _{u \in C}$ is larger than $0.5$ (that is, the average participation rate of users in community~$C$), for any community $C$. According to this criterion, we label $12,292$ elite Sybil users in total. Instead of binary classification, \textsc{ElsieDet} ranks elite Sybil users according to the \emph{Sybilness} score function (see Equation~\eqref{score}).

To carry out the ultimate validation on elite Sybil users detected from \textsc{ElsieDet}, we rely on human knowledge. In concrete, for each detected elite Sybil user, we manually categorize his or her reviews into two types, suspicious reviews and normal reviews, by inspecting Sybil campaign time intervals. The manual check then considers the following three criteria (by rule of thumb): \emph{(i) this user is involved in vast Sybil campaigns}; \emph{(ii) the intent of suspicious reviews is aligned with that of Sybil campaigns}. For example, in order to boost reputation in a Sybil campaign, the suspicious reviews should be $5$-star; \emph{(iii) suspicious reviews set apart from normal reviews in terms of spatio-temporal characteristics}. If a user satisfies all three criteria, we validate that he or she is an elite Sybil user. We emphasize that the criteria of manual validation are stricter than holding the two conditions carried out by \textsc{ElsieDet}.

Finally, of all the top $1,000$ suspicious elite Sybil users that our system flags, through manual validation, we conclude that $938$ are indeed elite Sybil users, which leads to a precision rate of $93.8\%$. We also randomly sampled $1,000$ flagged users to generalize the validation results, which also leads to a high precision rate of $90.7\%$.

\subsection{System Performance}\label{subsec:system performance}
We evaluate the efficiency of \textsc{ElsieDet} in a server with Intel CPU E3-1220 v3 @ 3.10GHz and 16G memory. Since \textsc{ElsieDet} has to compute potential collusion set and the pairwise similarity between potential collusive users to construct Sybil social links, this step would be the bottleneck of efficiency. Instead, we implement a parallel program for this step based on the observation that the computation for each user is independent. Finally, we implement a single-threaded program to complete following steps. Specially, for Dianping's dataset, the step of computing the pairwise similarity takes approximately $110$ minutes and the remaining steps take approximately $22$ minutes.

\section{Measurement and Analysis}\label{measurement}

In this section, we analyze the behavior of elite Sybil users and communities.
First, we compare the behavior patterns among benign users, regular Sybil users, and elite Sybil users.
We then discuss the relation between Sybil communities and elite Sybil users and review manipulation in chain stores.
We study reviews, not belonging to Sybil tasks, posted by elite Sybil users to speculate their strategies to camouflage fake reviews.
Finally, we demonstrate two temporal dynamics characterized by user posting period and Sybil campaign duration.

\subsection{Comparison with Regular Sybil Users}\label{ssec:micro look}

\begin{figure}[t]
  \begin{center}
  \includegraphics[width = 0.45\textwidth]{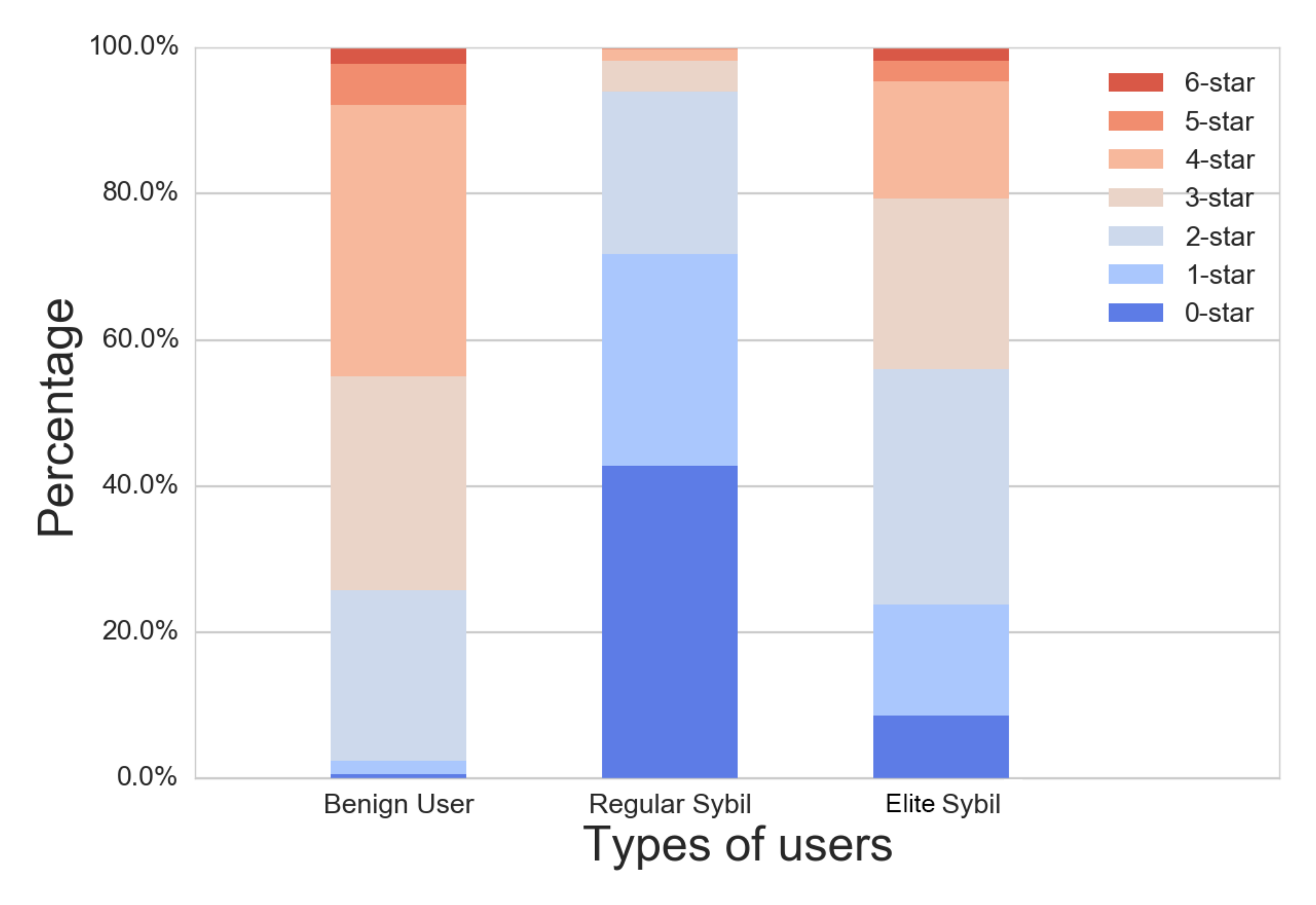}
  \vspace{-0.3cm}
  \caption{Comparison on distribution of user-level star ratings}
  \label{fig:percentage-user-rank}
  \vspace{-0.2cm}
  \end{center}
\end{figure}

Here, we try to explore the distribution of different types of user levels. Figure~\ref{fig:percentage-user-rank} shows that the distribution of levels of users is unevenly distributed for each type of users. As we can see from Figure~\ref{fig:percentage-user-rank}, we find that the distribution of levels of benign users is almost symmetrically bell-shaped, centered at 3-star. In contrast, the levels of regular Sybil users are heavily skewed toward low-level. Based on our results, we observe that the levels of elite Sybil users detected are biased more toward high-level than those of regular Sybil users.

Comparing elite Sybil users with regular Sybil users at the micro level, we show that elite Sybil users post more fake reviews, are more spread out temporally, and have fewer reviews filtered by Dianping. Figure~\ref{fig:Comparison senior Sybil regular Sybil} compares the behavioral patterns among elite Sybil users, regular Sybil users, and benign users on Dianping.

Figure~\ref{fig:comparison fake reviews} plots the CDF of the number of users in terms of the number of suspicious reviews posted. As can be seen in Figure~\ref{fig:comparison fake reviews}, elite Sybil users post the most suspicious reviews among all. This demonstrates that elite Sybil users cater to market demand due to their potential larger impact on Dianping ranking and higher prices for the customers. For the regular Sybil users, their strategy is frequently changing their low-level accounts to evade the detection since it is easy to apply or buy with a low cost for low-level accounts.

Figure~\ref{fig:percentage fake reviews} plots the CDF of the number of users in terms of the percentage of fake reviews posted. As we can see, fake reviews are significantly more often generated by regular Sybil users than by elite Sybil users, which echoes our definition that elite Sybil users post massive reviews not belonging to Sybil tasks (smoke-screening) to mimic genuine users. Surprisingly, the distribution of regular Sybil users roughly follows the Pareto principle (also known as the 80-20 rule) that more than $60\%$ of all the reviews posted by $20\%$ of regular Sybil users are fake. In contrast, as we can see from Figure~\ref{fig:percentage fake reviews}, we show that only $20\%$ of all the reviews posted by more than $80\%$ of elite Sybil users are fake, recognizing that the principle also applies in reverse.

Figure~\ref{fig:comparison filtered reviews} plots the CDF of the number of users in terms of their percentage of filtered reviews. As can be seen from Figure~\ref{fig:comparison filtered reviews}, we show that the percentage of filtered reviews of regular Sybil users significantly outnumbers that of benign users with respect to the same cumulative probability. To be specific, we see that $80\%$ of Sybil users (resp. benign users) have more than $90\%$ (resp. less than $20\%$) of reviews filtered. This user-level observation is consistent with the community-level results shown in Figure~\ref{fig:CDF-filtered-reviews-seed}. In addition, elite Sybil users have fewer reviews filtered by Dianping mainly because a large portion of their reviews are not assigned to any task.


\subsection{Community Structure}
Understanding the behaviors of elite Sybil users is important to reveal the characteristics of the (quasi) permanent workforce of Sybil organizations on Dianping. Looking at the macro level, communities of elite Sybil users form large-scale sparsely knit networks and their graph density is much lower.


\begin{figure}[h!]
  \begin{center}
  \includegraphics[width = 0.42\textwidth]{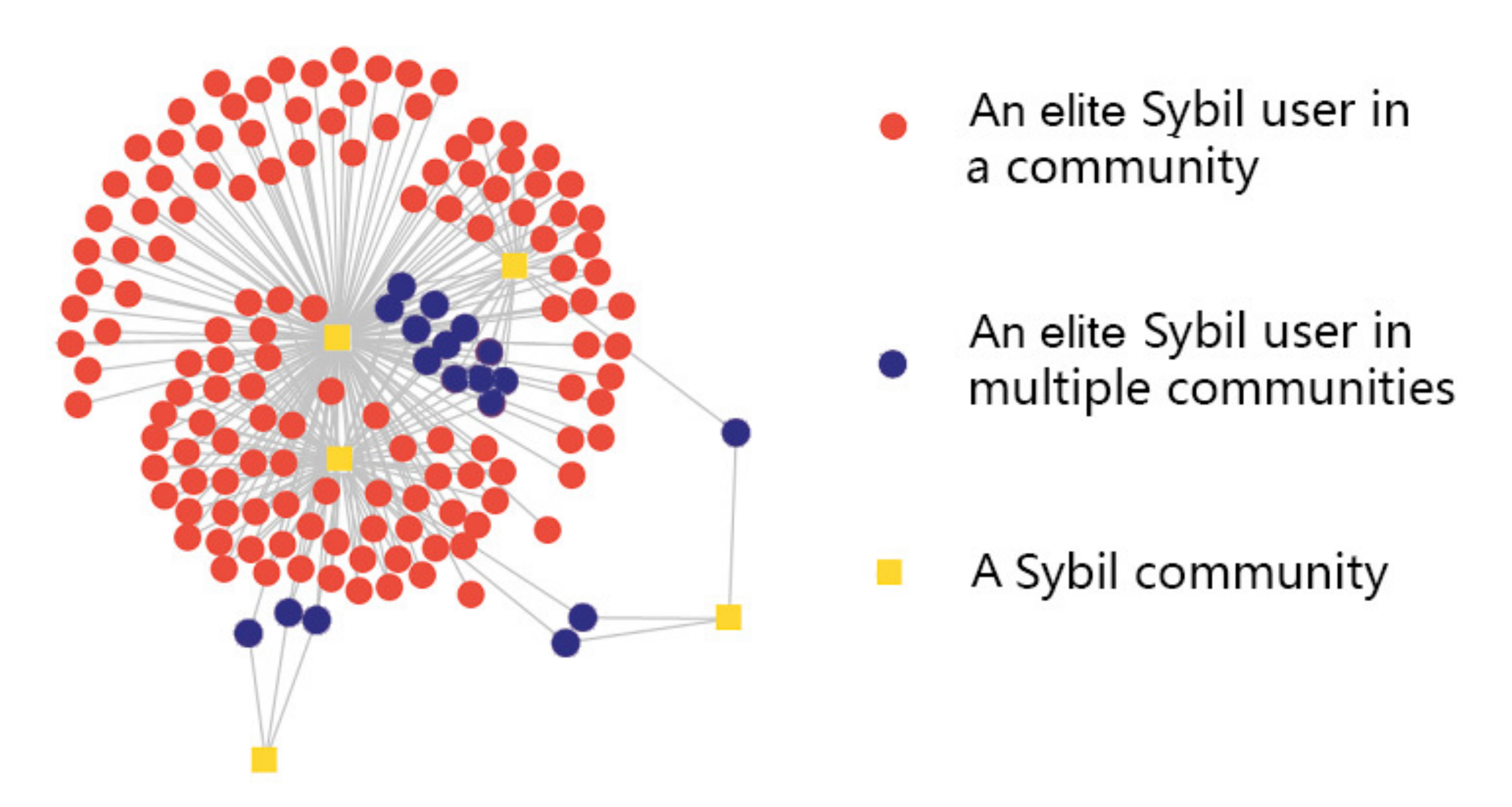}
   \vspace{-0.3cm}
  \caption{Relation between elite Sybil users and communities
  }
  \label{fig:relation}
 \vspace{-0.3cm}
  \end{center}
\end{figure}

Figure~\ref{fig:relation} shows an example of an induced network structure of elite Sybil users. 
In the figure, a dot represents an elite Sybil user, a square represents a Sybil community, an edge between a dot and a square represents that an elite Sybil user belongs to a community, and a red (resp. blue) dot shows that an elite Sybil user belongs to a single community (resp. multiple communities).
As can be seen, we observe that many elite Sybil users are correspondingly connected to a single community, forming a large-scale sparsely knit network. We also show that some elite Sybil users appear in multiple communities. Ranked by \emph{Sybilness}, we pick up the top $1,000$ elite Sybil users out of all $12,292$ users in our collection. There are $824$ elite Sybil users who participated in a single community, $160$ who participated in two communities, and $16$ who participated in at least three communities. Not surprisingly, we clearly show that these elite Sybil users are sparsely connected and their graph density is much lower than that of regular Sybil users.

\subsection{Review Manipulation for Chain Stores}

Recent research from Harvard~\cite{harvard2016fake} pointed out that it is less likely for chain stores to hire Sybil accounts to generate favorable reviews.
Chain stores tend to depend heavily on various forms of promotion and branding to establish their reputation. This is because chains receive less benefit from reviews, and they may also incur a larger cost if they are caught posting fake reviews, destroying their brand image. However, our research contradicts this statement. We find that a series of chain stores leverage Sybil organizations to post fake reviews to manipulate their online ratings.

To be more specific, of all $566$ Sybil communities in our dataset, it is observed that $12.37\%$ of Sybil communities post fake reviews for chain stores listed on Dianping. The number of chain stores involved varies from $2$ to $11$. One possible explanation is that the chain stores hired the same Sybil agent, who recruited the same Sybil community for Sybil campaigns.

\begin{figure}[t]
  \begin{center}
  \includegraphics[width = 0.42\textwidth]{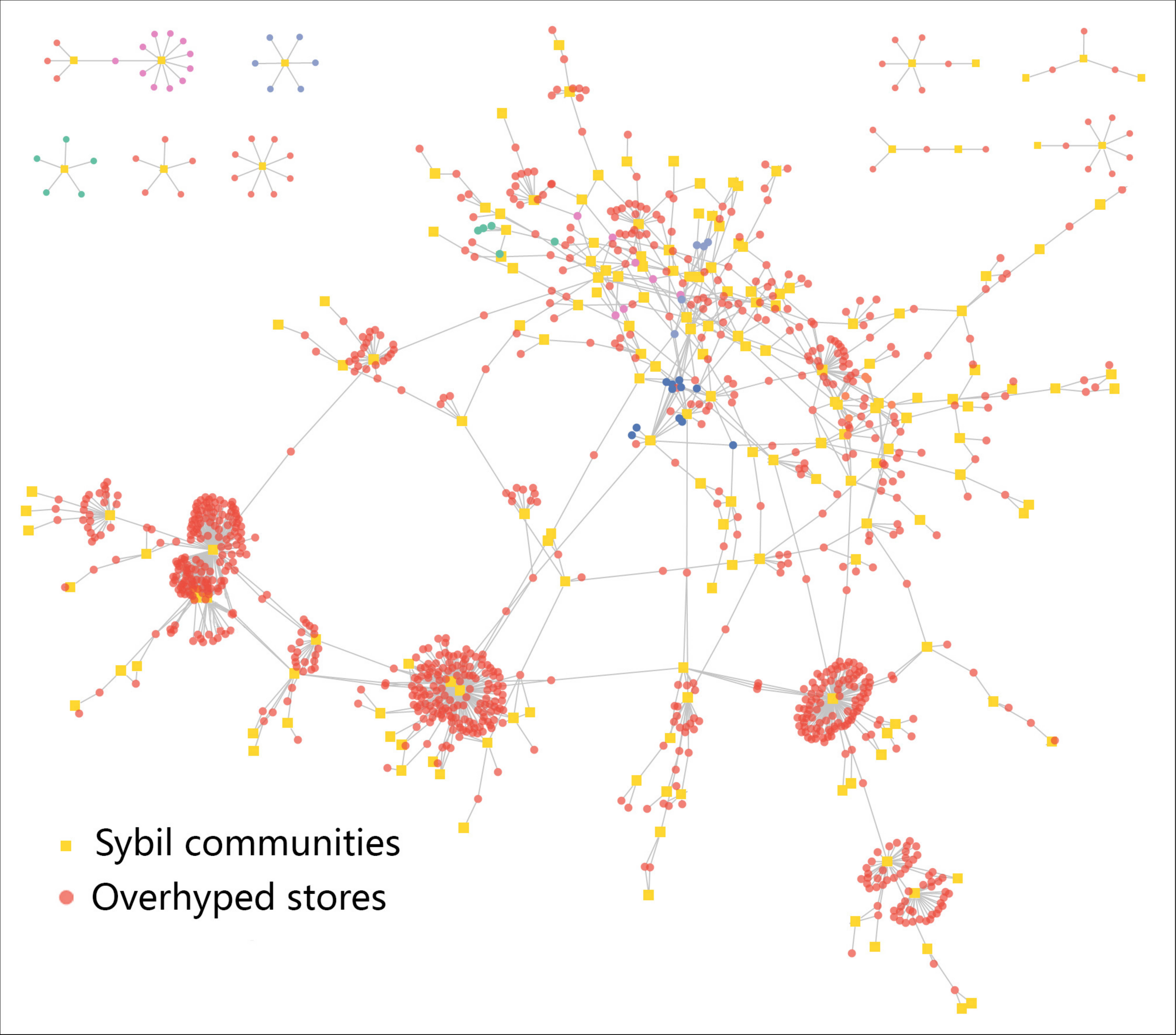}
  \caption{Relation between Sybil communities and the overhyped stores} 
  \label{fig:Community-shop}
 \vspace{-0.6cm}
  \end{center}
\end{figure}

Figure~\ref{fig:Community-shop} shows the main part of the entire network structure of Sybil communities and overhyped stores, pruned by a small portion of tiny networks. 
In the figure, a yellow square represents a Sybil community, a red dot represents an overhyped store, and an edge between a yellow and a red dot represents that a Sybil community connects to an overhyped store.
As can be seen, almost all Sybil communities act as central nodes. This indicates that these Sybil communities not only launch campaigns for a single store, but also provide various services for a huge number of overhyped stores who are connected by the network. Furthermore, some overhyped stores connect to multiple communities, which indicates that they have employed Sybil communities more than once (A case study is detailed in Section~\ref{ssec: multi-communities}.). We also label chain stores that have at least five branches with different colors other than red. These chains are connected to the same communities, respectively, possibly sharing similar reviews and having the same goal.

\subsection{Early Alerts for Sybil Campaigns}\label{ssec:online-detect}

In this subsection, we will show that it is feasible to uncover Sybil campaigns through monitoring our detected elite Sybil users. In particular, by continually monitoring the collusive behaviors of elite Sybil users, the social network operator can determine whether a Sybil campaign has been launched at the earliest stage, which serves as an early alert for a Sybil campaign. 

\noindent \textbf{Detecting Sybil campaigns via monitoring elite Sybil users.}
Our goal is to detect the presence of a Sybil campaign at the early stage based on identifying elite Sybil users via continually monitoring all elite Sybil users. To do this, we simply apply 7-day slide windows along the timeline to each store so as to detect campaigns. The rule of determining a Sybil campaign is more than a predetermined threshold number (\emph{e.g.}, 7 in our experiment) of reviews that the elite Sybil users posted at the same store within a 7-day slide window. Our heuristic is that, in the non-campaign period, the elite Sybil users normally post reviews at different stores in similar ways as innocent users due to their different living habits, walking routines, or shopping preferences. However, only within the campaign period, the elite Sybil users collusively post reviews at the same stores to fulfill the Sybil campaign tasks. The evaluation results show that by scanning the activities of elite Sybil users during the entire campaign period, approximately $90.40\%$ campaigns can be determined. This indicates that the campaign determination rule holds for almost all the Sybil campaigns.

\noindent \textbf{Determining Sybil campaigns at the early stage.}
An interesting question is whether we can determine a Sybil campaign at the early stage. The benefits of early detection is that it can give a competitive advantage for the system operator to take countermeasures against Sybil campaigns. We run the campaign window determination algorithm by using the first $1/4$, $1/3$, and $1/2$ of the entire campaign period. The evaluation results show that $56.77\%$, $63.08\%$, and $75.14\%$ of campaigns can be successfully detected correspondingly. Since the average Sybil campaign period is $68$ days in our experiments, it indicates that more than $50\%$ of Sybil campaigns can be determined within the first two weeks by only observing activities of elite Sybil users, thereby triggering lightning strikes on Sybil campaigns.


\subsection{Temporal Dynamics}
We demonstrate two temporal dynamics characterized by user posting period and Sybil campaign duration.

\begin{figure}[t]
  \begin{center}
  \includegraphics[width = 0.4\textwidth]{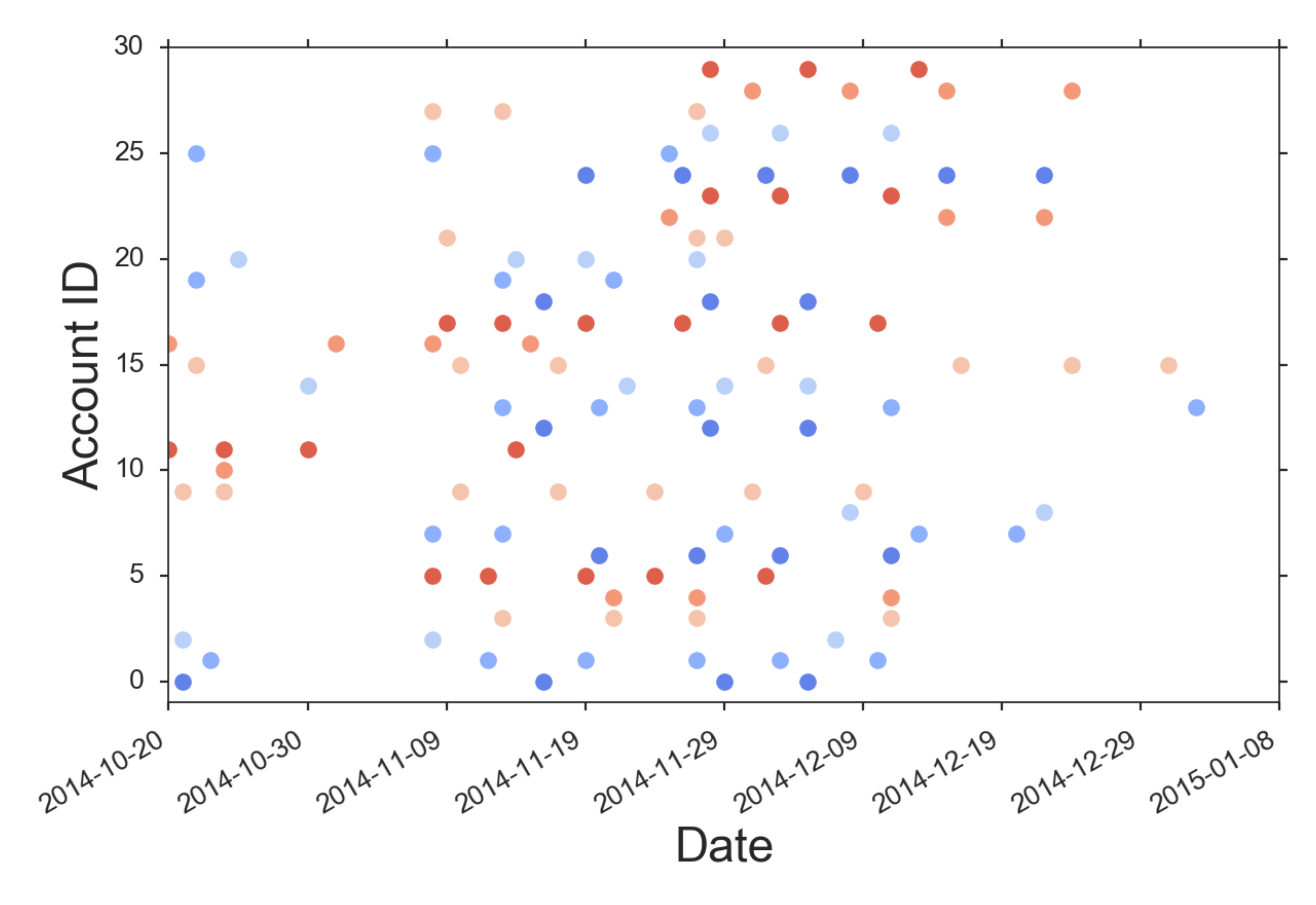}
   \vspace{-0.3cm}
  \caption{Reviews posted by Community $4559$ in Store $4112200$
  }
  \label{fig:case-study-period}
\vspace{-0.6cm}
  \end{center}
\end{figure}

\noindent \textbf{User posting period.} Figure~\ref{fig:case-study-period} shows that elite Sybil users in Community $4559$ repeatedly post fake reviews in Store $4112200$. In the Figure, the x-axis shows the time when an account posts a review, and the y-axis is the account's ID. A dot $(x,y)$ in the figure represents that an account with ID $y$ posts a review at time $x$. We use staggered colors to encode reviews posted by different users. As we can see from Figure~\ref{fig:case-study-period}, $33$ users in Community $4559$ posted $127$ reviews within a period of two months. Posting reviews by these users is much denser than by benign users. Apart from posting reviews within a short time period, these elite Sybil users also deliberately manipulate posting time of reviews. For example, some elite Sybil users even periodically (every week/month) post fake reviews. We emphasize that manipulation of posting temporal dynamics is key to orchestrating the evasive strategy.

\begin{figure}[t]
  \begin{center}
  \includegraphics[width = 0.45\textwidth]{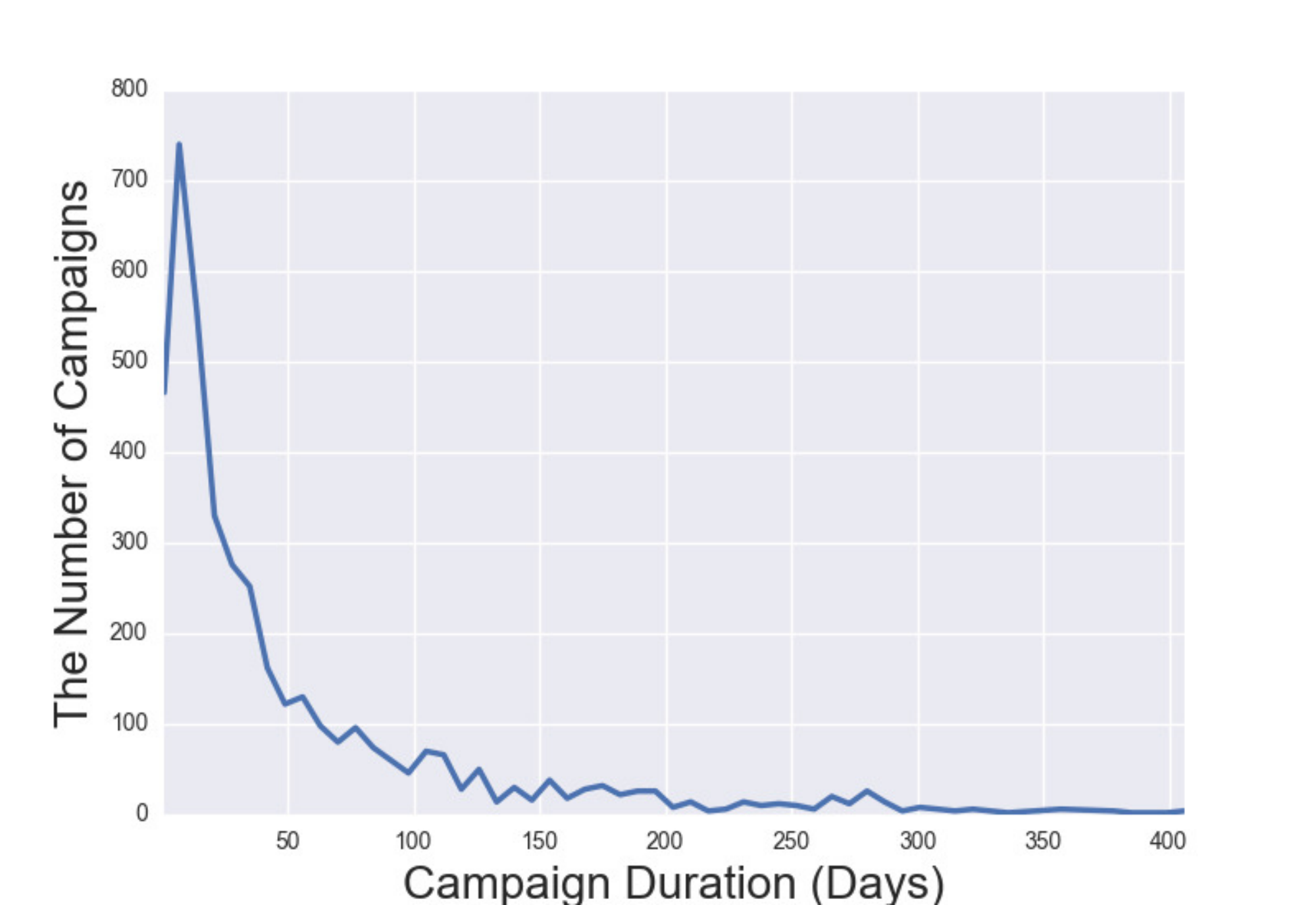}
   \vspace{-0.5cm}
  \caption{The distribution of number of campaigns across campaign duration}
  \label{fig:campaign-time}
 \vspace{-0.8cm}
  \end{center}
\end{figure}

\noindent \textbf{Sybil campaign duration.} By applying the campaign window detection algorithm, we finally obtain $4,162$ Sybil campaigns. Figure~\ref{fig:campaign-time} shows the distribution of number of campaigns across campaign duration. As we can see from Figure~\ref{fig:campaign-time}, the distribution is unimodal with a sudden spike at 7 days for the x-axis, echoing our 7-day slide windows selected; then largely monotonically decreasing beyond $50$ days. More remarkably, we observe there are $466$ 1-day ephemeral Sybil campaigns as shown by the y-intercept of Figure~\ref{fig:campaign-time}. In these campaigns, Sybil communities generally complete a task fleetingly.

In this section, we will detail a case study of Sybil communities and campaigns and illustrate various strategies to evade the Dianping's Sybil detection system.

\subsection{Sybil Communities and Sybil Campaigns} \label{ssec: multi-communities}
\begin{figure}[t]
  \begin{center}
  \includegraphics[width = 0.45\textwidth]{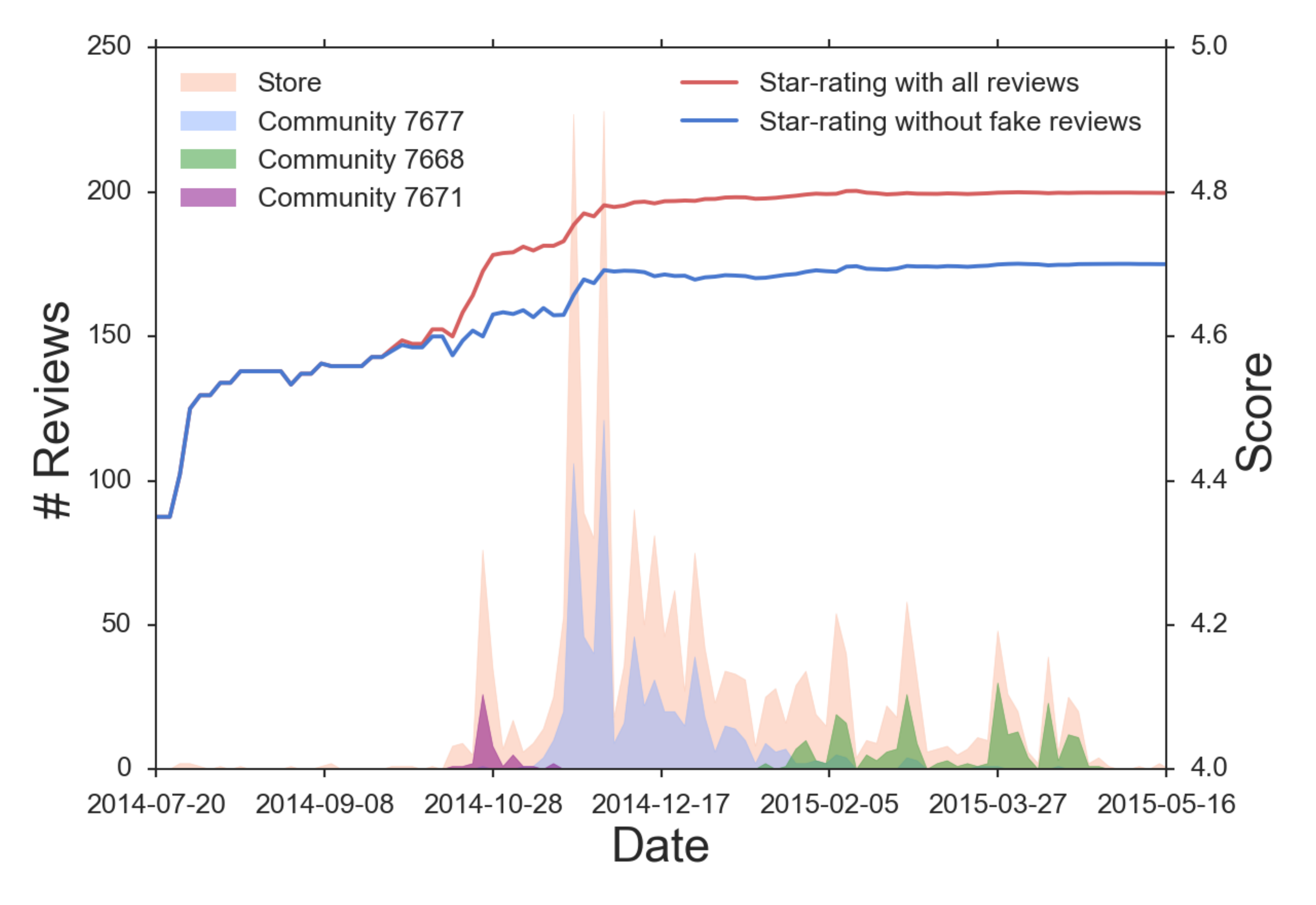}
   \vspace{-0.3cm}
  \caption{Variation of star ratings and the number of reviews of a hotel}
  \label{fig:case-study-hotel}
  \vspace{-0.5cm}
  \end{center}
\end{figure}

Recall in Figure~\ref{fig:Community-shop}, we show a part of stores employ several Sybil communities to increase their star ratings. Here, we zoom in and show the first case study that is about a hotel employing three different Sybil communities to post fake reviews. Figure~\ref{fig:case-study-hotel} shows that the variation of the star rating and the number of reviews change over time. Orange represents aggregate reviews of a hotel; blue, purple, and green represent reviews coming from three respective Sybil communities, respectively. The red line denotes the star rating of a hotel and the blue line denotes the star rating without detected fake reviews.

As we can see from Figure~\ref{fig:case-study-hotel}, many spikes occurred, generated by three Sybil communities, always correspond to the spike of the total number of reviews. This indicates that these fake reviews causing sudden spikes are taken into effect to raise the star rating of the hotel. In addition, as pointed out from Figure~\ref{fig:case-study-hotel}, red and blue lines are overlapping before the first spike; the red line then increases sharply afterwards but the blue line maintains a moderate growth. This indicates that these fake reviews posted by Sybil communities do have an impact on distorting the online rating. Figure~\ref{fig:case-study-hotel} also implies that Community $7677$ commits the largest-scale fake reviews and contributes most to increasing the star rating. However, Community $7668$ launches a fairly long-term Sybil organization but takes a very ``moderate'' gain on the star rating. This is perhaps because the hotel has had accumulated a significant number of reviews previously. Another possible reason is that the secret ranking algorithm adopted by Dianping does not merely depend on the average rating of a store. Features, such as the number of reviews and the number of page views, are another factors to determine the rank of a store. Hence, although these reviews do not have a discernible impact on the average star rating, they may also affect ranking results on Dianping.

%
%
%


\subsection{Evading Dianping's Sybil Detection System}\label{ssec:case-evading2}

In this case study, we present three examples of elite Sybil users in the same community to attempt to illuminate the evasive strategy taken by elite Sybil users. We also compare the results processed by Dianping's filtering system with ours.

\begin{figure}[htb]
	\centering
    \subfigure[No reviews filtered \label{fig:case-study-senior(a)}]{\includegraphics[width=0.33\textwidth]{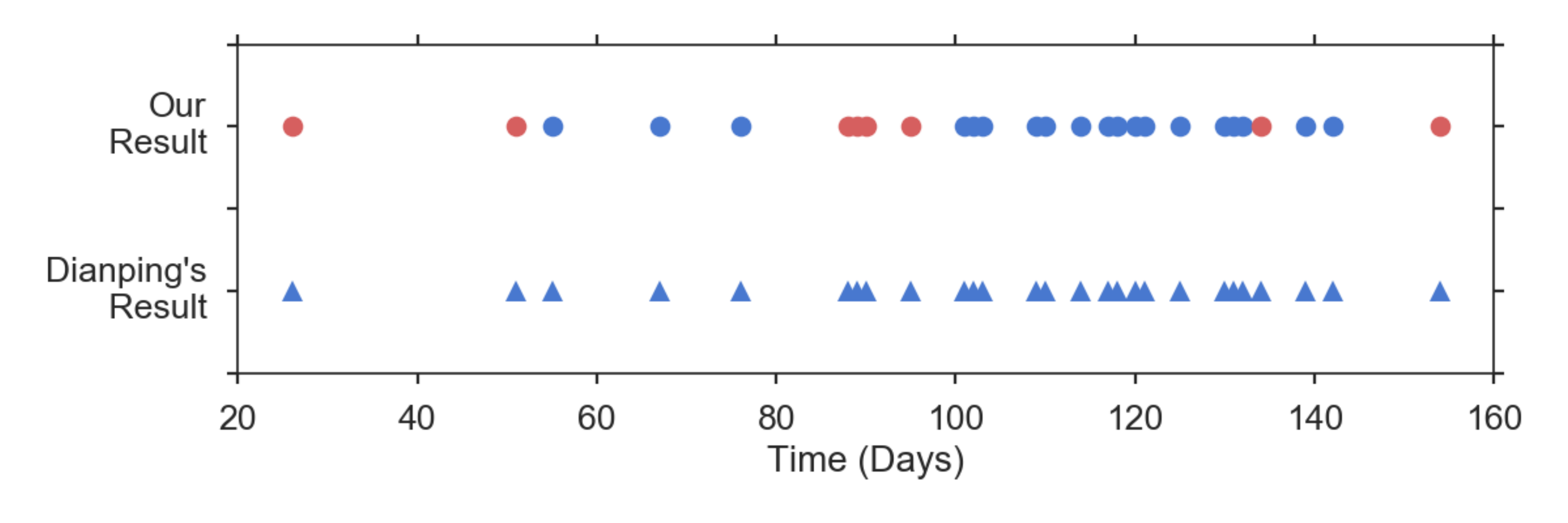}}
    \subfigure[Partial reviews filtered \label{fig:case-study-senior(b)}]{\includegraphics[width=0.33\textwidth]{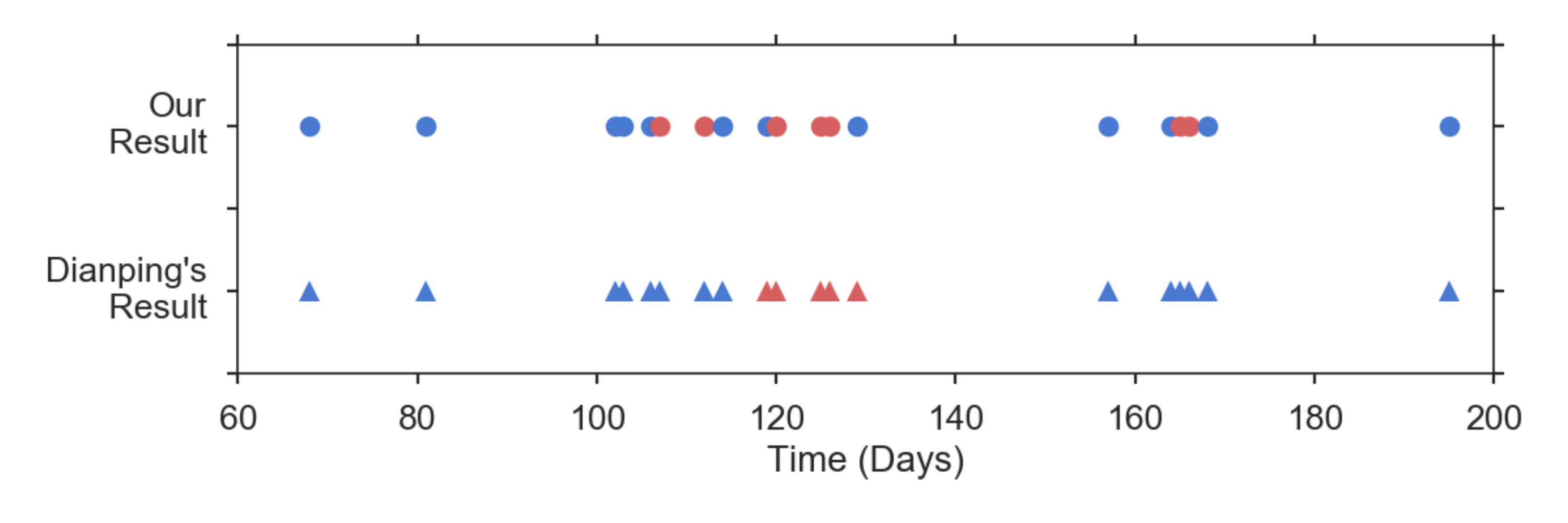}}					    
    \subfigure[All reviews filtered \label{fig:case-study-senior(c)}]{\includegraphics[width=0.33\textwidth]{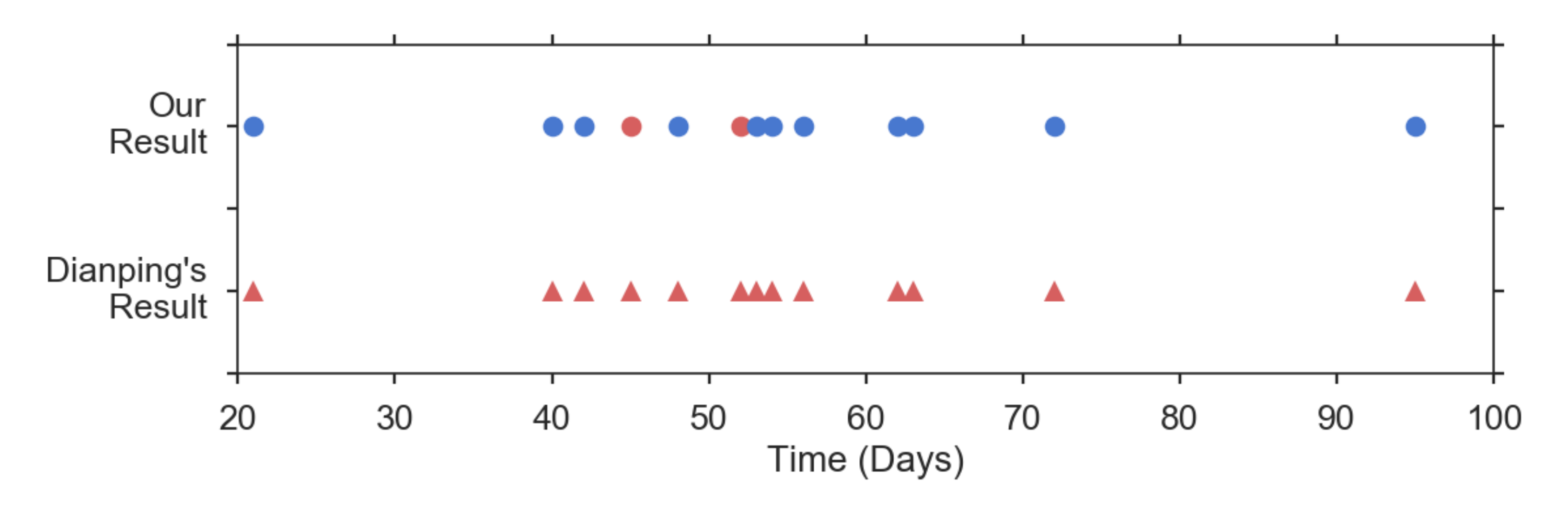}}
    \vspace{-0.2cm}
     \caption{Different detection results for elite Sybil users}
     \label{fig:case-study-seniorsybil}
     \vspace{-0.4cm}
\end{figure}

As shown in Figure~\ref{fig:case-study-seniorsybil}, each subfigure corresponds to the reviews posted by each elite Sybil user. Each dot represents a review posted according to the timeline; the upper (resp. lower) dotted line of reviews represents the posting timeline generated \textsc{ElsieDet} (resp. Dianping filtering system). In specific, each blue (resp. red) dot represents real (resp. fake) reviews labeled by \textsc{ElsieDet}. Each blue (resp. red) triangle represents existing (resp.  filtered) reviews according to Dianping. Through these three examples, first, we can analyze the evasive strategy taken by elite Sybil users. Elite Sybil users will post massive reviews to camouflage fake reviews and this strategy can evade most aggregate behavioral-based clustering approaches that rely on computing similarity of user activity. Second, these three users appear in the same community. They write fake reviews in similar stores and share with the similar behavioral patterns in the way to post reviews. However, there is one distinct difference from Dianping filtering system. For a given user, in Figure~\ref{fig:case-study-senior(a)}, we can see that no reviews have been filtered; in Figure~\ref{fig:case-study-senior(b)}, partial reviews have been filtered; in Figure~\ref{fig:case-study-senior(c)}, Sybil users are extremely sensitive to Dianping filtering system as all reviews have been filtered. This is perhaps because of his/her long negative-credit history.

\textit{In summary, we feel that Dianping filtering system is largely working on detecting regular Sybils, 
as shown in Figure~\ref{fig:case-study-senior(b)}.
We feel that Dianping is being fairly opaque about its filtering system as most of the real reviews of an elite Sybil user have also been falsely filtered due to its high false alarm rate, as shown in Figure~\ref{fig:case-study-senior(c)}. Although our dataset is moderate in size compared with the Dianping database, it is large enough to allow us to gain meaningful insights and identifying factors that impact the results and limitations of conventional Sybil detection systems.}

\vspace{-0.1cm}

\section{Discussion and Limitations}\label{ssec:application}
In this section, we discuss the potential application of \textsc{ElsieDet} and the limitations of the paper.

\subsection{Application of \textsc{ElsieDet}}
We here show how \textsc{ElsieDet} can be integrated to the existing Dianping's Sybil system to enhance its tolerance of elite Sybil attacks.



\noindent {\bf{Mitigating Sybil attacks by changing the weight of reviews with respect to Sybilness.}} The ultimate goal of a Sybil campaign is to manipulate the ratings of stores by generating massive fake reviews and ratings. To mitigate the negative impact of Sybil attacks on stores' ranking, a potential approach is to tune the weights of reviews of the suspicious users according to their \textit{Sybilness}. By assigning a lower weight to a highly suspicious user, it will significantly increase the difficulty of the Sybil organizations to manipulate the ratings and help alleviate the human labor required to verify massive number of users reported.

\noindent {\bf{Monitoring the top elite Sybil users to predict Sybil campaigns.}} Detecting Sybil campaigns is critical for Dianping to limit the impact of Sybil attacks. In Section~\ref{ssec:online-detect}, we have pointed out that we can monitor elite Sybil users and exploit their group actions to identify the Sybil campaign in a real-time fashion. Note that, considering millions of stores and users, only monitoring a small set of suspicious users can significantly save the efforts and resources of the social network operators. 

\subsection{Limitations}
First, 
although our detection system has strictly focused on Dianping, our results are applicable to a wider range of URSNs or any social media that relies on user-contributed comments. Examples include E-commerce (Amazon, Ebay, BizRate), movie-rating platforms (IMDB, Netflix, Douban), traveling services (TripAdvisor), and multi-agent systems (Advogato). In specific, in 2012, Yelp profile pages featured ``consumer alerts'' on several sneaky businesses which got caught red-handed trying to buy reviews, crafted by Yelp ``elite'' users, for these businesses~\cite{NYTyelp}. TripAdvisor has also put up similar warning notices. These examples may have specific detection systems, and we leave their design and evaluation to future work. Second, 
we acknowledge if a Sybil community can minimize the involvement in multiple campaigns, it would be very likely to boost the chance to evade the detection; however, recruiting high-cost elite Sybil users to participate in limited Sybil campaigns contradicts the economic basis. Third, we do not study the relationships among reviewers on Dianping. For example, a reviewer can make friends and keep a friend list on Dianping. A reviewer can send a flower to another reviewer in order to present a sense of complement to the reviewer who posts a nice review. We think these social links among reviewers are weak, extraneous for characterizing elite Sybil users on Dianping. Instead, we exploit user-community as a zoom lens to take a particular micro-macro analysis of elite Sybil users without using any user profile information.


\section{Related Work}\label{relatedwork}

In this section, we survey the methodology used in previous research from four categories: graph-based approaches, feature-based approaches, aggregate behavioral-based clustering approaches, and crowdsourcing-based approaches. We review each of these approaches as follows.

\noindent \textbf{Graph-based approaches.} Graph-based detection views accounts as nodes and social links between accounts as edges. For example, Liu \emph{et al.}~\cite{liu2015exploiting} considered the dynamic change in the social graph. Much prior work~\cite{mohaisen2010measuring,cao2012aiding,gong2014sybilbelief} holds the assumption that in a social graph, there exist a limited number of attack edges connecting between benign and Sybil users.  
The key insights behind this is that it becomes difficult for attackers to set up links to real users, and strong trusts are lacking in real OSNs, such as RenRen~\cite{yang2014uncovering} 
and Facebook~\cite{ikram2017measuring, DeCristofaro:2014:PLU:2663716.2663729, boshmaf2011socialbot, bilge2009all}.
Souche~\cite{xie2012innocent} and Anti-Reconnaissance~\cite{paradise2014anti} also rely on the assumption that social network structure alone separates real users from Sybil users. Unfortunately, this was proven unrealistic since real users refuse to interact with unknown accounts~\cite{ stringhini2010detecting}.
Recent research~\cite{NDSS2015integro} relaxes these assumptions and takes a combined approach that first leverages victim prediction to weigh the graph and upper bound the aggregate weight on attack edges; then it performs a short random walk on the weighted graph and distributes manually-set scores to classify users. 
However, We argue that these methods do not hold on URSNs and the nodes in URSNs do not show a tight connectivity as those in general OSNs, which renders the social network graph-connectivity-based Sybil detection approaches less effective in URSNs.

\noindent \textbf{Feature-based approaches.} The advantage of behavioral patterns is that these can be easily encoded in features and adopted with machine learning techniques to learn the signature of user profiles and user-level activities. Different classes of features are commonly employed to capture orthogonal dimensions of users' behaviors~\cite{rahman2014turning,stein2011facebook,egele2013compa,song2015crowdtarget,34click2013usenix,Li2015Analyzing}. Other work~\cite{thomas2011suspended,ramachandran2007filtering,thomas2011design} considers the associated content information, such as reviews context, wall posts, hashtags, and URLs,
to filter Sybil users. Specifically, the Facebook immune system~\cite{stein2011facebook} detects Sybil users based on features characterized from user profiles and activities.
COMPA~\cite{egele2013compa} is designed to uncover compromised accounts via sudden change alerts according to the behavioral patterns of users. 
In addition to user profile, Song \emph{et al.}~\cite{song2015crowdtarget} proposed a target-based detection on Twitter approach which bases on features of retweets.
However, feature-based approaches are relatively easy to circumvent by adversarial attacks~\cite{zhang2016adversarial,biggio2014security,wang2014sparse,bruckner2012static}.  Further work will also be needed to detect sophisticated strategies exhibiting a mixture of realistic and Sybil users features.

\noindent \textbf{Aggregate behavioral-based clustering approaches.} Recently, rather than classifying single users, much work~\cite{34click2013usenix,35www2013copycatch,CCS14syn,CCS14twitter,www12fim, gao2010detecting} focuses on detecting clusters of users. Specifically, CopyCatch~\cite{35www2013copycatch} 
and SynchroTrap~\cite{CCS14syn}, implementing mixed approaches, score comparatively low false positive rates with respect to single feature-based approaches. For Dianping, the elite Sybil users, however, write elaborate reviews by mimicking the real reviews and intentionally manipulate the review temporal patterns within a Sybil campaign, so as to change the behavior features to bypass detection. 

\noindent \textbf{Crowdsourcing-based approaches.}
Wang \emph{et al.}~\cite{wang2013social} tested the efficacy of crowdsourcing (such as leveraging humans, both expert annotators, and workers hired online), at detecting Sybil accounts simply from user profiles. The authors observed that the detection rate for hired workers drops off over time, although majority voting can compensate for the loss. However, two drawbacks undermine the feasibility of this approach: (i) This solution might not be cost effective for large-scale networks, such as Facebook and Dianping; (ii) exposing personal information to external workers raises privacy issue~\cite{elovici2014ethical}. We observe that some recent work discusses how to identify the regular Sybil users in URSNs (\emph{e.g.}, Yelp and Dianping) by exploiting crowdsourcing-based approaches~\cite{lee2013crowdturfers, rahman2014turning, song2015crowdtarget}, or model-based detection~\cite{libimodal} that limits their broad applicability. Most recent work leverages Recurrent Neural Networks (RNNs) to automate the generation of synthetic Yelp reviews~\cite{CCS17crowdturfing}. However, we emphasize that \textsc{ElsieDet} is immune to the AI attack for two reasons: (i) \textsc{ElsieDet} does not accommodate any contextual features that RNN-based attack is centered around. (ii) The attack dataset used in~\cite{CCS17crowdturfing} does not take in any human-crafted fake reviews, which presumes that the proposed defense~\cite{CCS17crowdturfing} cannot well identify the fake reviews written by elite Sybil users defined in our paper.
We believe that our research is the first to define, characterize, and perform a large-scale empirical measurement study toward the elite Sybil attack in URSNs. We thus 
hope that our results may serve as a supplement to other traditional Sybil detection schemes and shed light on the novel Sybil detection system for uncovering other evolved Sybil users.

\vspace{-0.2cm}

\section{Conclusion}\label{conclusion}
This paper illuminates the threat of large-scale Sybil activities in User-Review Social Networks. We first demonstrated that Sybil organizations of Dianping utilize a hybrid cascading hierarchy to orchestrate campaigns. An in-depth analysis of elite Sybil users leads us to several important conclusions: 
elite Sybil users are more spread out temporally, craft better-edited contents, but have fewer reviews filtered. We showed that most Sybil campaigns can be determined within the first two weeks by only monitoring detected elite Sybil users. Strikingly, we also showed that a series of chains leverage Sybil organizations to distort the online rating, rendering previous research outdated. We emphasize that sophisticated manipulation of temporal patterns is key to orchestrating the evasive strategy. Finally, we demonstrated that \textsc{ElsieDet} is both highly effective and scalable as a standalone system.

Although our study and experiments focus on Dianping, we believe that the anti-Sybil defense as examined in this paper provides an opportunity for all URSNs to stop the spread of elite Sybil users in a way that has never been visible on Dianping or other social networks like it.

\section*{Acknowledgment}
This work was supported in part by the National Science Foundation of China, under Grants 71671114, 61672350, and U1405251. Corresponding author: Haojin Zhu.



%
\bibliographystyle{IEEEtranS}
\bibliography{sigproc-bp1}

\end{document}